\documentclass[lettersize,journal]{IEEEtran}

\usepackage{microtype}
\usepackage{subfig, siunitx, adjustbox}
\usepackage{booktabs} 

\usepackage{amsfonts}
\usepackage{array}
\usepackage{textcomp}
\usepackage{stfloats}
\usepackage{url}
\usepackage{verbatim}
\usepackage{graphicx}
\usepackage{cite}

\usepackage{amsmath}
\usepackage{amssymb}
\usepackage{mathtools}
\usepackage{amsthm}
\usepackage{adjustbox}
\usepackage{algpseudocode}
\usepackage{algorithm}
\usepackage{wrapfig}

\usepackage{multirow}
\usepackage{makecell}
\usepackage{booktabs}
\usepackage{colortbl}
\usepackage{xcolor}

\makeatletter
\AfterEndEnvironment{algorithm}{\let\@algcomment\relax}
\AtEndEnvironment{algorithm}{\kern2pt\hrule\relax\vskip3pt\@algcomment}
\let\@algcomment\relax
\newcommand\algcomment[1]{\def\@algcomment{\footnotesize#1}}

\renewcommand\fs@ruled{\def\@fs@cfont{\bfseries}\let\@fs@capt\floatc@ruled
  \def\@fs@pre{\hrule height.8pt depth0pt \kern2pt}%
  \def\@fs@post{}%
  \def\@fs@mid{\kern2pt\hrule\kern2pt}%
  \let\@fs@iftopcapt\iftrue}
\makeatother

\theoremstyle{plain}
\newtheorem{theorem}{Theorem}

\theoremstyle{definition}
\newtheorem{definition}{Definition}

\theoremstyle{remark}
\newtheorem{remark}{Remark}

\usepackage{hyperref}

\usepackage[compress, numbers]{natbib}

\usepackage{titlesec}
\titleformat{\paragraph}[runin]
  {\normalfont\normalsize\bfseries}{}{0em}{}
\titlespacing{\paragraph}
  {0pt}{0.5ex plus 0.5ex minus .2ex}{1em}

\begin{document}

\title{
    Noise Variance Optimization in Differential Privacy: \\
    A Game-Theoretic Approach Through Per-Instance Differential Privacy
}

\author{Sehyun Ryu*,
        Jonggyu Jang*,~\IEEEmembership{Member,~IEEE,}
        Hyun Jong Yang,~\IEEEmembership{Member,~IEEE}
    \thanks{
    S. Ryu, J. Jang, and H. J. Yang are with the Department of Electrical Engineering at Pohang University of Science and Technology (POSTECH) (Email: \{sh.ryu, jgjang, hyunyang\}@postech.ac.kr).
    The corresponding author is Hyun Jong Yang.
    }
    \thanks{$(^*)$: These authors are equally contributed.}
}



\maketitle


\begin{abstract}

The concept of differential privacy (DP) can quantitatively measure privacy loss by observing the changes in the distribution caused by the inclusion of individuals in the target dataset. 
The DP, which is generally used as a constraint, has been prominent in safeguarding datasets in machine learning in industry giants like Apple and Google. 
A common methodology for guaranteeing DP is incorporating appropriate noise into query outputs, thereby establishing statistical defense systems against privacy attacks such as membership inference and linkage attacks.
However, especially for small datasets, existing DP mechanisms occasionally add excessive amount of noise to query output,  thereby discarding data utility.
This is because the traditional DP computes privacy loss based on the worst-case scenario, i.e., statistical outliers. 
In this work, to tackle this challenge, we utilize per-instance DP (pDP) as a constraint, measuring privacy loss for each data instance and optimizing noise tailored to individual instances. 
In a nutshell, we propose a per-instance noise variance optimization (NVO) game, framed as a common interest sequential game, and show that the Nash equilibrium (NE) points of it inherently guarantee pDP for all data instances. 
Through extensive experiments, our proposed pDP algorithm demonstrated an average performance improvement of up to 99.53 \% compared to the conventional DP algorithm in terms of KL divergence.

\end{abstract}


\section{Introduction}
\label{sec:introduction}

\begin{figure}[t]
\vspace{0pt}
    \centering
    \includegraphics[draft=false, width= 0.99 \linewidth]{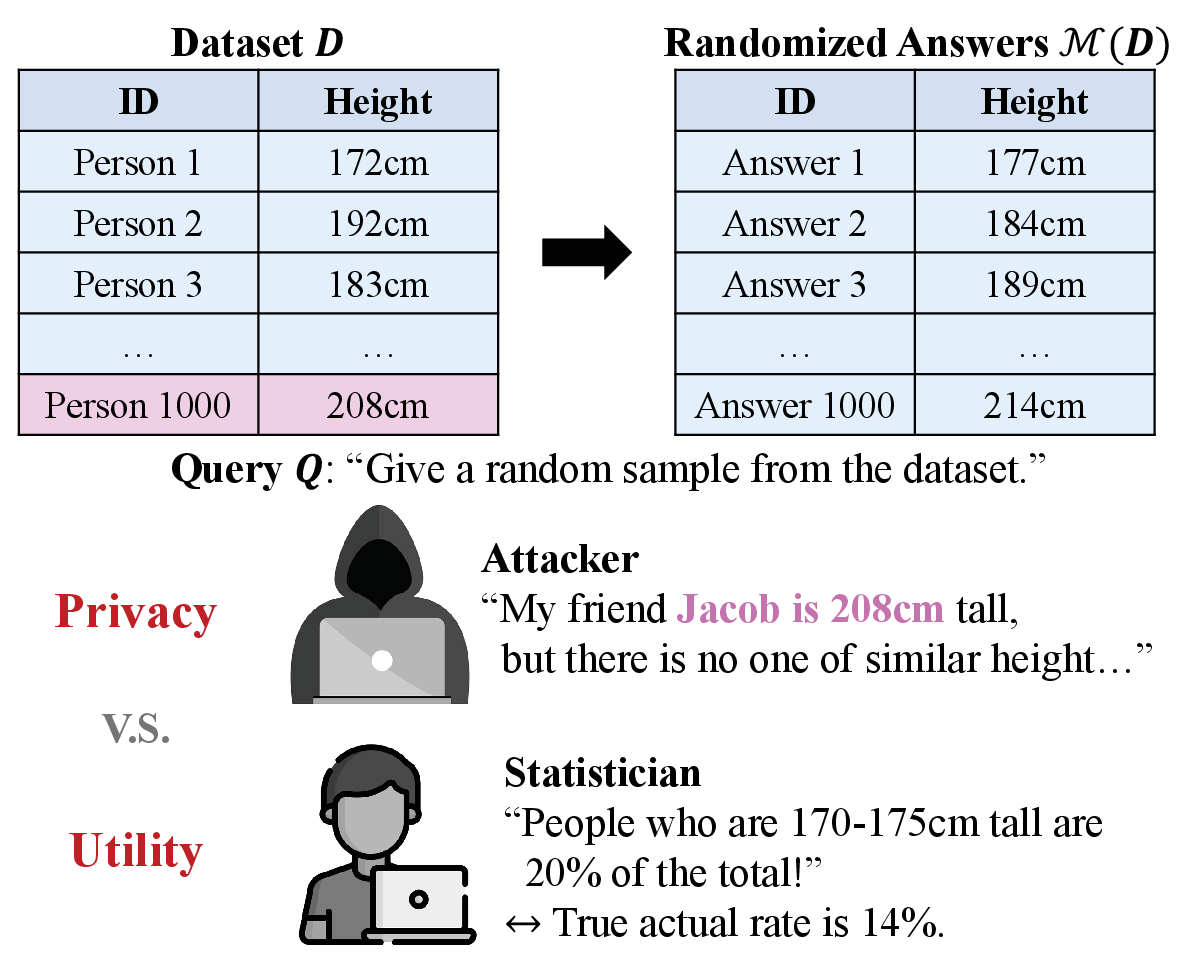}
\vspace{-8pt}
    \caption{The essence of DP's data security lies in injecting noise into query outputs. The noise level has to be delicately balanced, considering the tradeoff between privacy and utility. From above, the attacker tries to identify if Jacob is in the dataset based on his known height of 208cm, but privacy is protected by noise, making it impossible. Meanwhile, the statistician struggles to analyze the feature of the dataset accurately due to noisy responses, resulting in utility loss.}
    \label{fig:concept}
    \vspace{-7pt}
\end{figure}

Recently, the surge in machine learning has not only spotlighted the importance of statistical datasets but has also intensified the focus on privacy protections. 
Meanwhile, the innovative concept of differential privacy (DP) has emerged as a key solution for quantitatively measuring privacy risks~\cite{DWORK2006}. 
As depicted in Fig. \ref{fig:concept}, the DP, a method that balances the utility of data with individual privacy by injecting controlled randomness into datasets, ensures that the privacy of individual data points is preserved by making it \textit{mathematically indiscernible} whether any specific individual data is included or excluded.
In the figure, from a dataset $\mathcal{D}$, the data agent answers a query $Q(\mathcal{D})$. If there is no randomness in the query output, one potential attacker may access the private information via membership inference~\cite{SHOKRI2017} or data linkage attacks~\cite{NARAYNAN2007}.
In DP, the basic concept is forwarding randomized query output $Q(\mathcal{M}(\mathcal{D}))$ or $\mathcal{M}(Q(\mathcal{D}))$, where $\mathcal{M}$ is a randomized mechanism, thereby making the query output mathematically indiscernible. 


In today's digital landscape, DP has been a promising technology as a strong safeguard against the looming threats of privacy violations, where the most common approach is simply adding noise to query outputs, i.e., \textit{additive noise mechanism.}
Specifically, compared to its rivals like homomorphic encryption and federated learning \cite{LEE2022, WU2023}, the additive noise DP mechanisms are computationally simple and generally applicable. 
By virtue of its simplicity and straightforwardness, DP has been widely employed by industry giants such as Apple \cite{TANG2017} Google \cite{ERLINGSSON2014}, and Microsoft \cite{JUSTICIA2022} for data protection. 

\paragraph{Limitations of DP.} 

Although the DP mechanisms have been found in our daily lives, they sometimes add an excessive amount of noise to the query output, thereby making the datasets' utility almost statistically useless. 
To introduce this limitation, we bring the definition of the $\epsilon$-DP, the basic concept of DP, where the $\epsilon$ is a privacy parameter. 

\begin{definition}[$\epsilon$-DP]
\label{def:epsilon-dp}
    A randomized mechanism with domain $\mathbb{R}^d$, denoted by $\mathcal{M}$, has a range $\mathcal{R}(\mathcal{M})$. The mechanism $\mathcal{M}$ is $\epsilon$-differential private, if for all dataset $\mathcal{D}, \mathcal{D}' \in \mathbb{R}^d$ such that $||\mathcal{D}-\mathcal{D}'||_{1} \leq 1$:
    \begin{equation}\label{eq:dp}
        \bigg|\ln\frac{\Pr[\mathcal{M}(\mathcal{D}) \in S]}{\Pr[\mathcal{M}(\mathcal{D}') \in S]}\bigg| \leq \epsilon, ~ \forall S \subseteq \mathcal{R}(\mathcal{M}).
    \end{equation}
\end{definition}

\noindent In Definition \ref{def:epsilon-dp}, the smaller $\epsilon$, the stronger the privacy guarantees in DP.
More importantly, the $\epsilon$-DP is broadly defined across arbitrary datasets $\mathcal{D}\in\mathbb{R}^{d}$ and data instances. 
That is, the traditional DP focuses on designing mechanism $\mathcal{M}$ that satisfies the condition $\eqref{eq:dp}$ for all statistical datasets $\mathcal{D}$, i.e., the mechanism $\mathcal{M}$ is designed for \textit{the worst-case scenario} of all the statistical datasets.
In general, the worst-case scenario indicates that $n-1$ data instances are the same, while one outlier has totally different information. 
Because this outlier can be placed anywhere, the conventional DP mechanisms, such as Logistic and Gaussian mechanisms \cite{CHEN2024}, reasonably add \textit{excessive} amount of \textit{identical} noise to the query outputs.

\begin{figure*}[t]
\vspace{0pt}
    \centering
    \includegraphics[draft=false, width= 0.78 \linewidth]{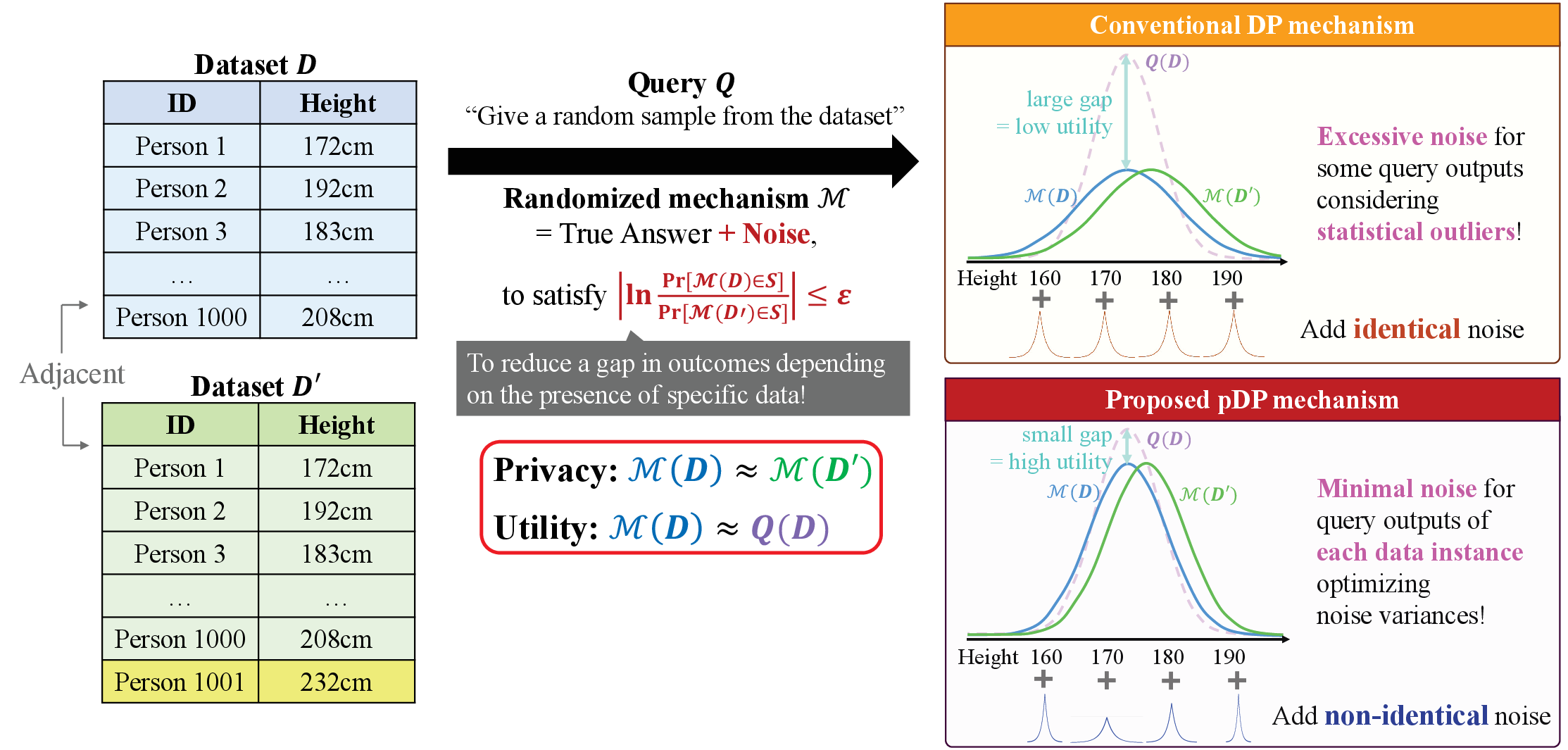}
    \caption{This figure illustrates the concept of DP guarantees. Conventional DP mechanisms sample noise values from the same probability distribution for all query outputs. Our proposed pDP mechanism aims to maintain the same level of privacy loss while mimicking the probability distribution of original query outputs by adding noise with varying variances to each instance's query output.}
    \label{fig:example}
\end{figure*}

\paragraph{Per-Instance DP.}
Although adding \textit{identical} noise in the conventional mechanisms is a reasonable choice in DP; however, if we only consider a specific dataset $\mathcal{D}$, we can further enhance the utility of the dataset by \textit{adopting non-identical noise} while preserving privacy. 
The recently proposed concept, per-instance DP (pDP), gives us a motivation that the privacy loss of an instance in a fixed dataset can be measured.

\begin{definition}[$\epsilon$-pDP \citep{WANG2019}]
\label{def:epsilon_pdp}
A randomized mechanism, denoted by $\mathcal{M}$, has a range $\mathcal{R}(\mathcal{M})$. For \textcolor{red}{a fixed dataset $\mathcal{Z}$} and \textcolor{red}{a fixed data instance $z\in\mathcal{Z}$}, the mechanism $\mathcal{M}$ meets $\epsilon$-pDP, if the following condition holds:
\begin{equation}
\bigg|\ln\frac{\Pr[\mathcal{M}(\mathcal{Z}) \in S]}{\Pr[\mathcal{M}(\mathcal{Z} \textcolor{red}{\setminus \{z\}}) \in S]}\bigg| \leq \epsilon, ~ \forall S \subseteq \mathcal{R}(\mathcal{M}),
\end{equation}
where \textcolor{red}{red}-colored contents are different from Definition \ref{def:epsilon-dp}. 
We note that the pDP is defined for fixed dataset $\mathcal{Z}$ and data instance $z$; thus, the DP in \eqref{eq:dp} is the supremum of pDP for dataset $\mathcal{Z}$ and instance $z$. 
\end{definition}

Although no specific solution to design noise distribution tightly guaranteeing pDP, the authors of \cite{WANG2019} provide measuring different security levels for each data instance, defining privacy loss within the actual dataset and analyzing the variations. 
If a specific dataset is given, we no longer need to consider the worst-case for all arbitrary datasets. 
As depicted in Fig. \ref{fig:example}, applying identical noise to all query outputs for DP satisfaction is undoubtedly not the optimal solution.

\paragraph{In this paper,} we focus on designing appropriate noise distribution for each data instance for simultaneously satisfying pDP for all data instances.
Intuitively, according to the definition of the pDP, the rarer the data instance, the harder it is to guarantee statistical indistinguishability. 
In the upcoming sections, we aim to answer the following question:
\begin{center}
    \textit{\textbf{
    How can we optimize noise on a per-instance basis to satisfy pDP for a dataset while preserving the statistical features of the original data as much as possible?
 }}
\end{center}

In response to this question, our objective is to introduce a per-instance additive noise mechanism grounded in the principles of pDP. 
We propose solving the problem as a noise variance optimization (NVO) game and establish, through Theorem \ref{thm:NE_NVO}, that the Nash equilibrium (NE) of this game necessarily ensures pDP for all data instances.


\paragraph{Challenges.} 

Here, we introduce challenges on the horizon in optimizing the noise distribution to guarantee pDP. 
The main challenge is that ensuring pDP for a particular data instance is inherently dependent on the noise distribution of other data instances.
Thus, altering the noise distribution for a data instance presents a tangible risk: certain instances might become non-compliant for pDP as a consequence of such modifications.  
Conventional additive noise mechanisms can guarantee mathematically well-proven assurance; however, when introducing non-identical noises, establishing these guarantees becomes more difficult.
In summary, finding a balance between preserving the dataset's original statistical utility and ensuring $\epsilon$-pDP requires intricate adjustments to the noise distribution, a challenge amplified by the curse of \textit{\textbf{interdependency}}.

\paragraph{Contributions.}
We introduce an innovative approach to optimize non-identical noise distribution tailored to specific data instances. Our main contributions are three-fold:
\begin{itemize}
    \item We propose the NVO game designed to find suitable non-identical per-instance additive Laplace noises within a dataset. Within this game, every player (representing data instances) collaboratively/sequentially acts to guarantee $\epsilon$-pDP, all the while optimizing the utility of data statistics.
    \item  We prove that an NE strategy in the NVO game ensures $\epsilon$-pDP across all data instances.
    \item We simulate the best response dynamics (BRD) algorithm and the approximated enumeration (AE) algorithm as examples to obtain an NE strategy for the proposed NVO game. The proposed NVO game not only assures the same $\epsilon$-pDP as the commonly adopted Laplace mechanism but also demonstrates superiority in preserving statistical utility.
\end{itemize}

\section{Related works}

\subsection{Additive mechanisms for DP}
Traditional additive noise mechanisms offer straightforward and mathematically well-proven methods to ensure DP. 
Many attempts have been made to improve these traditional methods: \cite{GENG2014, GENG2015, GENG2019} have proposed additive staircase-like noise as a substitute for the Laplace distribution, aiming to optimize a given statistical utility function while guaranteeing $\epsilon$-DP.
In addition, the IBM DP library demonstrates efforts to enhance additive noise mechanisms by constraining the randomized output within a predefined range \citep{HOLOHAN2019}. 
In the context of optimizing the noise distribution, \cite{MIRONOV2017} has attempted to regulate additive noise to meet R\'{e}nyi DP criteria.
Similar concepts have been explored in parallel studies: sampling scenarios~\citep{GEUMLEK2017, GIRGIS2021} or deep learning~\citep{WANG2022, ZHU2020}.
While there are efforts to alter noise distributions, prior studies have primarily employed uniform noise across all data.
Such a method is not appropriate for guaranteeing tight pDP while preserving statistical utility through per-instance non-identical noise.

\subsection{Relation to \texorpdfstring{$(\epsilon,\delta)$}{}-DP}
The pioneer of DP~\cite{DWORK2014G}, defined $(\epsilon,\delta)$-DP, 
 demonstrating that the Gaussian mechanism can achieve this definition.
Since the advent of DPSGD~\cite{ABADI2016}, there has been a surge of applications in machine learning~\cite{DING2021, MOREAU2021, TRUEX2020}. It is worth noting that our method can be easily adapted to the widely-known relaxation of DP, -- namely ($\epsilon,\delta$)-pDP, with the per-instance Gaussian mechanism.
Furthermore, when the dataset displays ample diversity among its values and there are constraints on accessing external data, ($\epsilon,\delta$)-pDP becomes equivalent to ($\epsilon,\delta$)-DP.
Thus, our research on the pDP mechanism can be interpreted as striving to ensure greater effectiveness of DP in specific scenarios.

\subsection{Game theory}
We introduce the NVO game, designed to simplify identifying the best variance for additive noises, inspired by the familiar game-theoretic approach introduced by~\cite{NEUMANN1944}.
Within the context of the NVO game, we show that the NE points of the game ensure $\epsilon$-pDP.
However, identifying this NE point brings its own set of complexities.
To address this, we devise methods utilizing established algorithms to reach the NE point~\citep{TAYLOR1978, ZAMAN2018}.



\section{Preliminary}
In this section, we introduce the preliminary concepts underpinning this paper. 
For brevity, we use scalar-form data instances in the remainder of this paper. 
We note that the concepts presented here can be seamlessly applied to vector-form data instances as well.

\paragraph{Laplace Mechanism.}
In addition to the $\epsilon$-DP and $\epsilon$-pDP in Definitions \ref{def:epsilon-dp} and \ref{def:epsilon_pdp}, we begin with the definition of the Laplace mechanism, the most recognized method to guarantee $\epsilon$-DP.

\begin{definition}[Laplace mechanism]
Given any query function $f : \mathcal{X} \rightarrow \mathbb{R}$ with $\ell_1$ sensitivity of $\Delta f\in\mathbb{R}$, the Laplace mechanism is defined as:
\begin{equation}
  \mathcal{M}_{\textnormal{L}}(x,f(\cdot),\epsilon) = f(x) + y,
\end{equation}
where $y$ is a random number drawn from $\textnormal{Lap}(\Delta f / \epsilon)$ and $\mathcal{X}$ denotes the domain of variable $x$.
\end{definition}

The Laplace mechanism ensures $\epsilon$-DP by considering the worst-case scenario and adding identical noise to all data instances.
However, we see an opportunity to enhance the preservation of statistical features in the data by employing a customized per-instance noise variance within the Laplace distribution.

\paragraph{Random Sampling Query.}
In this study, we focus on the random sampling query, as detailed subsequently. 
\begin{definition}[Random sampling query]\label{def:rsq}
Given numeric dataset $\mathcal{D}$, the output of the random sampling query $q$ is a random variable following the probability distribution of a dataset, \textit{i.e.},
\begin{equation}
  q(\mathcal{D}) \sim \Pr(\mathcal{D}).  
\end{equation}
\end{definition}
This query captures the statistical characteristics of the dataset by directly retrieving an instance from it.

\begin{remark}[Extensibility of random sampling query\label{remark:ran_samp_query}]
    The random sampling query is a fundamental query that encompasses all possible statistical queries. This is because the random sampling query can capture the statistical distribution of a dataset. Thus, from the post-processing theorem, achieving pDP/DP for random sampling queries can guarantee pDP/DP for all statistical queries.
\end{remark}

\paragraph{Nash Equilibrium.}
The Nash equilibrium, a foundational concept in a game theory introduced by \cite{NASH1951}, denotes the ideal state of a game where every player makes their optimal decision based on the choices of their counterparts as below.

\begin{definition}[Nash equilibrium]
The Nash equilibrium is a profile of strategies $(s_{i}^{*}, s_{-i}^{*})$, such that each player's strategy is an optimal response to the other players' strategies:
$\Pi_{i}(s_{i}^{*},s_{-i}^{*}) \geq \Pi_{i}(s_{i},s_{-i}^{*}), ~ \forall i$
where $s_{-i}$ is the strategy profile of all players except for player $i$ and $\Pi_{i}(s)$ is a payoff function.
\end{definition}

To minimize complexity and structure the problem in a well-known format, our approach frames the challenge as a game to ensure pDP within a dataset and maximize the preservation of statistical features, ultimately reaching an NE point.

\begin{figure*}[t]
\vspace{-7pt}
    \centering
    \includegraphics[draft=false, width= 0.99 \linewidth]{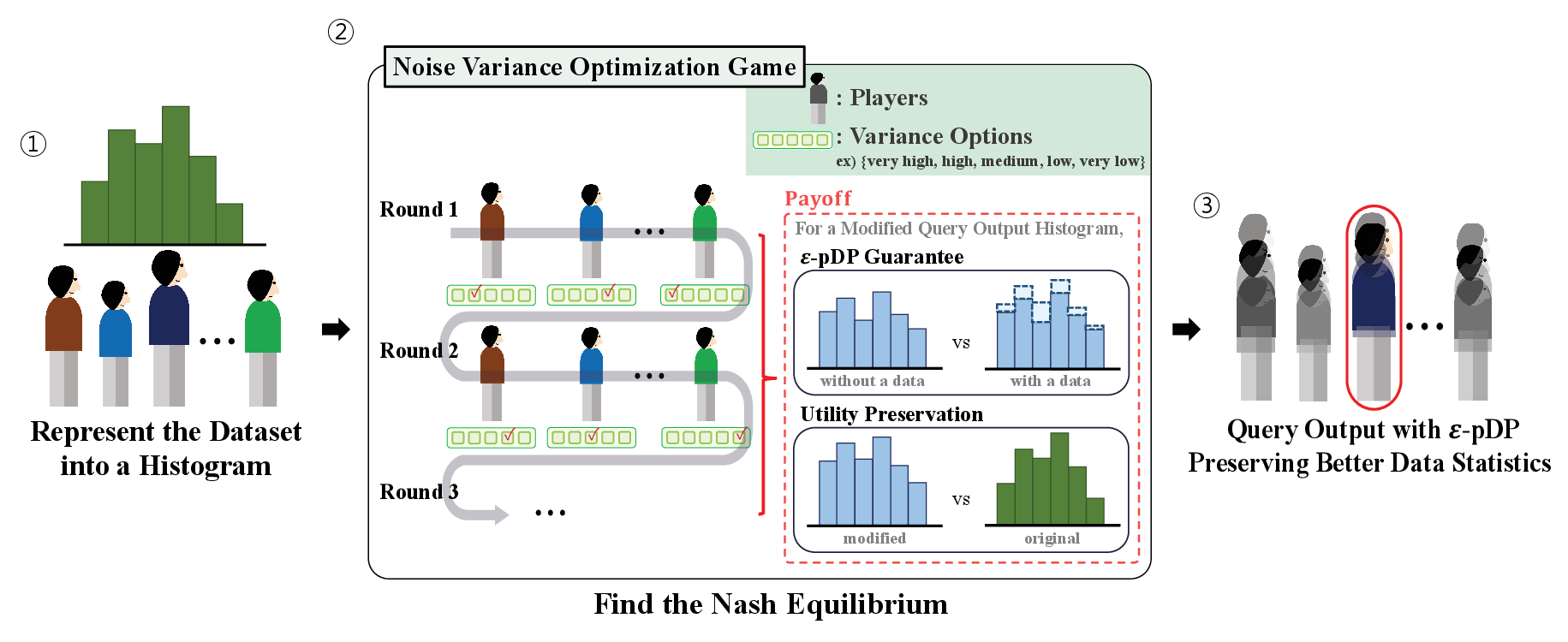}
\vspace{-7pt}
    \caption{
    The process involves determining the optimal combination of noise variances for the query outputs of individual data instances. In Step 1, we depict the dataset as a histogram through normalization and categorization. In Step 2, we aim to find an NE point for the NVO game, where multiple players iteratively update their variance parameters to ensure $\epsilon$-pDP while preserving statistical utility. 
    Once the prime set of noise variances is determined, Step 3 allows us to formulate queries that assure $\epsilon$-pDP by executing a random sampling query.}
    \label{fig:Workflow}
    \vspace{-7pt}
\end{figure*}

\section{Noise variance optimization game}
\label{SECTION:4}

In this section, our focus is to design a sequential/cooperative game that applies per-instance Laplace noises to the target dataset, ensuring $\epsilon$-pDP. 
We denote the target dataset by $\mathcal{D}$, and its data instances are represented by $d\in\mathbb{R}$.
We assume that the target dataset consists of real-valued data instances, \textit{e.g.}, a regression dataset. 
The problem we aim to solve using game theory can be defined as a constrained optimization problem:
\begin{align}
\min_\mathcal{M} & ~~ U(\mathcal{D}, \mathcal{M}(\mathcal{D})) \\
\textnormal{s.t.} & ~~ \left|\ln\frac{\Pr[\mathcal{M}(\mathcal{D}) \in S]}{\Pr[\mathcal{M}(\mathcal{D} \setminus \{d\}) \in S]}\right| \leq \epsilon,  \forall d \in \mathcal{D}, \forall S\in\mathcal{R}(\mathcal{M}), \nonumber
\end{align}
where the function $U$ represents an arbitrary utility function, quantifying the statistical difference between the original dataset $\mathcal{D}$ and the randomized dataset $\mathcal{M}(\mathcal{D})$.
In our work, \textit{KL divergence} is used as our utility function.
Solving the problem as a conventional optimization problem becomes challenging due to the non-differentiable nature of the constraint function, making it difficult to apply gradient-based methods, and the presence of intricate interdependencies.
To address this, we simplify the problem as selecting a variance value among possible options and elucidate it using familiar game theory principles.

An illustration of the proposed game design, including the preprocessing step, is depicted in Fig. \ref{fig:Workflow}.
Detailed explanations corresponding to this figure will be covered in the subsequent portions of this section.

\subsection{Preprocessing: A histogram representation of the dataset}

In the context of the NVO game, assessing the probability density function of the mechanism's output for every single point across all data instances is computationally challenging. 
Additionally, as highlighted by \cite{ABADI2016}, the most precise method for assessing privacy loss involves manual integration within specified intervals, rather than relying solely on theoretical boundaries.

\paragraph{Normalization.}

For the manual integration, we normalize the dataset into the $[0,1]$ range by min-max normalization. 
We conservatively opt to set our integration's target range to encompass $p$-percentile of the Laplace mechanism with the target $\epsilon$, which is defined by $\left( -(\Delta q/{\epsilon})\ln(2 - 2p), (\Delta q /\epsilon) \ln(2 - 2p) \right)$ if $p>0.5$\footnote{In our experiments, the value of $p$ is set to be 0.9.}, where $\Delta q$ denotes the sensitivity of random sampling query $q$. 
For brevity, we denote $(\Delta q /\epsilon) \ln(2 - 2p)$ as $\Delta_{(\epsilon,p)}$. 
Then, the min-max linear normalization is executed in the interval 
$[d_{\textnormal{min}}-\Delta_{(\epsilon,p)},d_{\textnormal{max}}+\Delta_{(\epsilon,p)}]$, where $d_{\textnormal{min}}=\min_{i}d_{i}$ and $d_{\textnormal{max}}=\max_{i}d_{i}$.

\paragraph{Categorization.}
After normalization, the continuous nature of the domain $S$ poses a unique challenge, differing from that in $\epsilon$-DP. It is necessary to verify the $\epsilon$-pDP condition for each instance and each point within set $S$, a task impractical to achieve in polynomial time. To address this issue, we make $K$ non-overlapping intervals into the range of the dataset and allocate each data instance $d\in\mathcal{D}$ into categories based on their corresponding intervals.
We set the $K$ uniformly divided non-overlapping intervals as follows: the $i$-th interval is $(\frac{i}{K},  \frac{i+1}{K} )$, where the representative value of each interval bin is the midpoint of the interval.
The set of the representative values is denoted as $\mathcal{K}$.
Through data categorization, the dataset can be represented in the form of a histogram. 
Segmenting data instances into distinct bins offers an advantage, as it allows certain data instances to inherently ensure a non-infinite $\epsilon$, while the original continuous dataset cannot\footnote{Once categorized, if a category contains three data points, the inherent $\epsilon$-pDP assurance for these instances is given by $\log{3/2}\approx0.4$.}.

\subsection{Definitions of players, strategies, and payoffs}

Here, we introduce the NVO game following data preprocessing. 
In this game, every data instance possesses data values within the range $[0,1]$.
The classes within our proposed NVO game include:  i) \textit{sequential game}, ii) \textit{fully-cooperative game}, iii) \textit{potential game}, and iv) \textit{common interest game}.

\paragraph{\textbf{Players.}}
In this study, each data instance, acting as a player, collaboratively and sequentially participates in the NVO game. 
The goal of them is to establish $\epsilon$-pDP (their respective payoffs) by designing their strategies (variance optimization), simultaneously maximizing the preservation of original statistical features.
We note that the player of the NVO game is represented by $I=\{1,2,\dots,\lvert \mathcal{D} \rvert\}$.

\paragraph{\textbf{Strategies.}}
With data instances cast as players, the strategy for player $i$ is defined by the additive noise applied to data $d_{i}$. 
Denoting the variance of additive noises to the $i$-th data instance $d_{i}$ as $b_{i}$, the action of the player $i$ is written by $b_{i}$.
That is, from the random sampling query in Definition \ref{def:rsq}, the per-instance Laplace mechanism is expressed as
\begin{equation}
  \mathcal{M}(d_i) = d_{i} + y_i,
\end{equation}
where $y_{i}$ is a random variable drawn form $\textnormal{Lap}(b_{i})$. 
Typically, games with strategy sets of uncountable infinity might not always have an NE solution.
As a result, we confine these strategy sets to a discrete domain.
In other words, the added variance is chosen from a discrete set $\mathcal{V} = \{ v_{1}, v_{2}, \dots, v_{n} \}$.

\paragraph{\textbf{Payoffs.}}
In the context of the NVO game, the payoff should induce the player to primarily uphold $\epsilon$-pDP and secondarily preserve the dataset's statistical features.
In this domain, there is a trade-off between data statistics and privacy. Emphasizing robust privacy can reduce query output quality, while maximizing the preservation of statistical features might compromise privacy. The goal is to optimize the preservation of statistical features without violating the $\epsilon$-pDP constraint, achievable by carefully adjusting noise variance on query results.
With this in mind, we express the overall payoff $P(\mathcal{M}, \mathcal{D})$ as a composite of two objectives: privacy assurance $P_{\textnormal{E}}(\mathcal{M},\mathcal{D})$, and minimizing utility that measures the statistical difference between the original dataset and the randomized dataset $P_{\textnormal{U}}(\mathcal{M},\mathcal{D})$.
We note that the players in the proposed NVO game cooperate to benefit from the shared payoffs.
Simply, the NVO game is characterized as a kind of common interest game.

\paragraph{Privacy assurance payoff.}

The payoff related to privacy assurance, denoted as $P_{\textnormal{E}}$, functions as an indicator of how effectively 
$\epsilon$-pDP is met for a dataset, assessed from the perspective of pDP.
Let us define $p_{\epsilon, i}$ as an indicator for representing whether the $i$-th data instance's pDP is satisfied or not, \textit{i.e.,}
\begin{equation}
    p_{\epsilon, i}(\mathcal{M},\mathcal{D}) = 
    \begin{cases} 
        1, & \textnormal{if $d\in\mathcal{D}$ satisfies the $\epsilon$-pDP condition.} \\
        0, & \textnormal{otherwise.}
    \end{cases}
\end{equation}
Then, the privacy assurance payoff $P_{\textnormal{E}}$ is defined as the number of data instances satisfying the $\epsilon$-pDP condition as
\begin{equation}\label{eqn:P_E}
    P_{\textnormal{E}}(\mathcal{M},\mathcal{D}) = \sum_{i=1}^{|\mathcal{D}|} p_{\epsilon, i}(\mathcal{M},\mathcal{D}).
\end{equation}

After establishing the privacy assurance payoff, two subsequent questions arise: one regarding the preservation of statistical features, and the other addressing the assurance of privacy at the NE point.
\begin{itemize}
    \item \textit{\textbf{Q1:} How can we define the payoff related to the preservation of statistical features?}
    \item \textit{\textbf{Q2:} Does the NVO game truly ensure $\epsilon$-pDP for all data instances using the privacy assurance payoff of (\ref{eqn:P_E})?}
\end{itemize}

\subsection{Utility preservation payoff \textit{(Answer to Q1)}}
In response to \textit{Q1}, we formulate the utility preservation payoff $P_{\textnormal{U}}(\mathcal{M}, \mathcal{D})$ to measure the statistical difference between the original dataset $\mathcal{D}$ and the randomized dataset $\mathcal{M}(\mathcal{D})$. Here, a higher value indicates a smaller difference. 
Furthermore, we ensure the utility preservation payoff does not compromise the assurance of $\epsilon$-pDP by scaling the targeted utility function $U$ into the range $[0,1]$. It is important to note that various statistical features can be incorporated into the utility function, encompassing metrics like the difference of $n$-th order momentum, Kullback-Leibler (KL) divergence, and Jensen-Shannon (JS) divergence, among others.

In this paper, for example, $U(q(\mathcal{D})||q(\mathcal{M}(\mathcal{D}))) = D_{\textnormal{KL}}(q(\mathcal{D})||q(\mathcal{M}(\mathcal{D})))$, i.e., we set the utility function by using KL divergence as in the following remark.
\begin{remark}[Examples of the utility function]
    For a dataset $\mathcal{D}$, the output probability distribution $q(\mathcal{D})$ of a query $q$, and a randomized function $\mathcal{M}$, the utility preservation payoff is defined as
    \begin{equation}\label{eqn:P_U}
    P_{\textnormal{U}}(\mathcal{M},\mathcal{D}) = 1 -\frac{D_{\textnormal{KL}}(q(\mathcal{D})||q(\mathcal{M}(\mathcal{D})))}{\log(K)} \in [0,1], 
    \end{equation}    
    where the utility function $D_{\textnormal{KL}}(q(\mathcal{D})||q(\mathcal{M}(\mathcal{D})))$ is bounded in $[0, \log(K)]$.
    The minus sign is used since the KL-divergence is a measure of information-theoretic distance between two probability distributions, where a smaller value indicates greater similarity between the distributions.
    The bound can be obtained by the fact that $\log(K) \ge \log\frac{q(\mathcal{D})}{q(\mathcal{M}(\mathcal{D}))}$.
\end{remark}



\subsection{Guarantee of the \texorpdfstring{$\epsilon$}{}-pDP \textit{(Answer to Q2)}}
Regarding the earlier question, we present proof illustrating that the NE strategy in the proposed NVO game consistently guarantees $\epsilon$-pDP for a dataset.

\begin{theorem}\label{thm:NE_NVO}
    Let us define the minimum variance in the set of possible action $\mathcal{V}$ as $b_\textnormal{min}\neq0$. Then, $\epsilon$-pDP for all data instances upholds if the following condition is satisfied:
    \begin{equation}\label{eq:NE_NVO}
        b_{\min} \ge \frac{1}{\log\left(1+ (|\mathcal{D}|-1) (\exp(\epsilon) -1)  \right)}.
    \end{equation}
\end{theorem}

In Theorem \ref{thm:NE_NVO}, we show that an NE point for the NVO game guarantees the $\epsilon$-pDP for all $d\in\mathcal{D}$ if the condition in (\ref{eq:NE_NVO}) holds. 
In the theorem, there always exists a value $b_\textnormal{min}$ that makes the NE point of the NVO game ensure the $\epsilon$-pDP for all $\epsilon\ge0$. 

\begin{remark}[Intuition of Theorem \ref{thm:NE_NVO}]
In Equation $\ref{eq:NE_NVO}$, if there are more data instances in the dataset, the influence of the individual data point diminishes, thereby allowing us to guarantee $\epsilon$-pDP easily. That is, if the value of $|\mathcal{D}|$ increases, we can guarantee $\epsilon$-pDP with smaller variance noise. On the other hand, if $\epsilon$ decreases to zero, query output with and without a data point should be statistically the same. Thus, the variance of the added noise becomes infinite, resembling a uniform query output distribution.
\end{remark}

\section{Algorithm finding the Nash equilibrium of the NVO game.}
\label{SECTION:5}

In this section, we delve into an algorithm designed to secure an NE strategy within the framework of the NVO game. We begin by showcasing the BRD algorithm, adapted specifically for this game. 

\begin{figure}[ht]
\begin{minipage}{.93\linewidth}
\begin{algorithm}[H]
\caption{BRD for NVO game}\label{algo:BRD}
\algcomment{\small{*\texttt{GET\_PAYOFF()} is a function of the proposed payoff by summing up Equations (\ref{eqn:P_E}) and (\ref{eqn:P_U})}.}
\textbf{Input} dataset $\mathcal{D} = \{d_i | i=1,...,m\}$, variance space $\mathcal{V} = \{v_{i}|i=1,...,n\}$, target epsilon value $\epsilon$

\begin{algorithmic}
   \State $i \gets 1$ and $p^{*} \gets 0$ \\
   \hfill $\triangleright$ Initialize data index and the best payoff
   \State V[$l$] $\gets$ randomly initiates $v \in \mathcal{V}$, \: for $l=1, \dots, |\mathcal{D}|$\\
   \hfill $\triangleright$ Initialize the best variance set
   \While{$p^{*}$ converges over the dataset} \label{euclidendwhile}
      \State $\mathrm{PAYOFF}[l]$ $\gets$ 0, \: for $l=1, \dots, |\mathcal{V}|$
      \For{$j \gets 1$ to $|\mathcal{V}|$}
        \State V\_temp $\gets$ V
        \State V\_temp[$i$] $\gets$ $v_{j}$
        \State $\mathrm{PAYOFF}[j]$ $\gets$ \texttt{GET\_PAYOFF}($\mathcal{D}$, V\_temp, $\epsilon$)\\
        \hfill $\triangleright$ Explore and store payoffs for all variance options
        \EndFor
    \EndWhile
    \State $p^{*} \gets \max$ $\mathrm{PAYOFF}$ and $j^{*} \gets \textnormal{argmax}$ $\mathrm{PAYOFF}$
    \State V[$i$] $\gets$ $v_{j^{*}}$\\
    \hfill $\triangleright$ Get the best variance for a current instance
    \State $i \gets (i+1) \bmod |\mathcal{D}|$
   \State \textbf{return} V \hfill $\triangleright$ Nash equilibrium
\end{algorithmic}
\end{algorithm}
\vspace{-7pt}
\end{minipage}
\end{figure}

\paragraph{\textbf{BRD algorithm.}} \:
The BRD algorithm is a concept in game theory where players, taking into account the current strategies of their opponents, opt for their most favorable response. 
During this iterative process, players sequentially decide on their best actions, which is presented in Alg.~\ref{algo:BRD}.
The choice of values within the variance space can be tailored to encompass all the possible noise variance values. 
As the cardinality of the variance space $\mathcal{V}$ expands, the algorithm's performance improves, but there is a significant increase in computational complexity. Therefore, it is crucial to define the variance set $\mathcal{V}$ considering the trade-off between computational complexity and utility.

\paragraph{\textbf{Common interest game \& potential game.}} \:
In a common interest game, participants share a unified payoff. A player's strategy change directly impacts both the potential function and their own payoff, classifying it inherently as a potential game. Essentially, every data point acts as a cooperative player aiming for a joint goal. By crafting a shared payoff to maximize and iteratively selecting the optimal noise for each data instance's output, achieving an NE is feasible.


\paragraph{\textbf{Convergence of BRD toward NE.}} \:
As shown by $\text{\cite{BOUCHER2017}}$, the BRD algorithm always converges into an NE point, if the target game belongs to one of the following games: potential games, weakly acyclic games, aggregative games, and quasi-acyclic games. 
As noted above, the NVO game is a potential game; thus, the BRD algorithm can obtain an NE point of the NVO game.

From the proof of Theorem 4.1 in the Appendix $\text{\ref{appendix:thm_proof}}$, when every variance option exceeds $b_{\min}$, there always exists a choice that consistently increases one pDP assurance at each round. Hence, guaranteeing $\epsilon$-pDP for all data instances is feasible after $|\mathcal{D}|$ rounds.
Intuitively, as players choose their optimal responses, either sequentially or simultaneously, the value of the potential function increases, reaching its maximum at some point. The strategy at this peak is the game's NE.

\section{Experiments}
\label{SECTION:6}
In this section, we evaluate the NE strategy of the NVO game.
Our primary focus is to observe if the proposed NVO game achieves superior preservation of the dataset's statistical features compared to the conventional Laplace mechanism while maintaining the same level of $\epsilon$-pDP.
To assess this, we conduct simulations on three publicly available datasets: 1) NBA player dataset~\cite{nba_player}, 2) personal income dataset~\cite{income}, and 3) credit profile dataset~\cite{credit_profile}.

\paragraph{\textbf{Experimental detail.}} \:
In our experiments, we configure the target $\epsilon$ values in $\{1, 2, 4, 8\}$. 
Additionally, we experimented on the NBA player's height dataset with extremely low target epsilon values of $\{0.1, 0.3\}$.
After normalization and discretization, the height and weight values belong to 101 categories, \textit{i.e.,} $K=101$.
For the action of the players,  variance set $\mathcal{V}$ is defined by $\{3 \times \Delta q / \epsilon, 2 \times \Delta q / \epsilon, \Delta q /\epsilon, 0.33 \times \Delta q / \epsilon, 0.2 \times \Delta q/\epsilon\}$.
From Theorem \ref{thm:NE_NVO}, $\epsilon$-pDP for the smallest $\epsilon$ is achievable with the configured variance set $\mathcal{V}$, since  $b_\textnormal{min} \approx 0.129$. That is, the NE points in the proposed NVO game ensure $\epsilon$-pDP.
For comparison, we additionally implemented an approximated enumeration (AE) algorithm based on the genetic algorithm, with excessive generations.
For more details, please refer to Appendix \ref{Appen:AE}.

\paragraph{\textbf{Metrics of data statistics.}}
In the experiments, we use the following metrics related to data statistics: 
\begin{itemize}
    \item \textbf{KL divergence}: 
    We measure the KL divergence between the probability distribution of the original dataset and the randomized dataset. 
    The lower KL divergence indicates better preservation of the information of the original dataset.
    \item \textbf{L1 loss of standard deviation (SD)}: 
    This metric measures the $\ell_1$ error between the standard deviation of the original dataset and the randomized dataset. 
    \item \textbf{Jaccard index}: The Jaccard index is calculated by representing values in a probability distribution exceeding a certain threshold as sets and then computing the intersection over the union (IoU) of the two sets. This measure quantifies the similarity between two probability distributions, where a value closer to 0 indicates dissimilarity, while a value closer to 1 signifies similarity between the distributions. We set the threshold to 0.001 to examine the probability distribution of query output during experiments and select significant values.
    \item \textbf{Cosine similarity} \citep{FURNAS1988}: The probability mass function can be viewed as a vector with probability values. We leverage the cosine similarity to measure the similarity between two probability distributions represented as vectors. 
\end{itemize}

\subsection{Experimental result 1: Height data}

\paragraph{\textbf{Dataset}.} In the NBA players dataset, we employ 1,307 data instances with the tuple of (height and weight) for players who

\begin{table*}[t]
\centering
\caption{Each algorithm's average computation time, KL divergence, L1 loss of SD, Jaccard index with a threshold 0.001, and cosine similarity are evaluated for the height data, for $\epsilon$ = 0.1, 0.3, 1, 2, 4, and 8. The modified query output distributions for all algorithms satisfy $\epsilon$-pDP.}
\label{tab:Result_of_Height}
    \adjustbox{width=0.7\linewidth}
    {
    \begin{tabular}{cccc ccc } 
        \toprule[1pt]
        Algorithm & $\epsilon$ & \makecell{Comp. time \\ \tiny{(minutes)}} $\downarrow$ & \makecell{KL \\ divergence} $\downarrow$ & \makecell{L1 SD \\ loss} $\downarrow$ & \makecell{Jaccard index  \\ \tiny{(threshold=0.001)}} $\uparrow$ & \makecell{Cosine \\ similarity } $\uparrow$ \\
        \cmidrule[0.7pt](l{1pt}r{1pt}){1-7}

        \multirow{2}{*}[-22pt]{BRD} &  0.1 & \underline{\textbf{8}} & \underline{0.0512} & \underline{0.0126} & \underline{0.4878} & \underline{0.9989} \\
         & 0.3 & \underline{\textbf{5}} & \underline{0.0393} & \underline{0.0122} & \underline{0.5882} & \underline{0.9994} \\
         & 1.0 & \underline{\textbf{4}} & \underline{\textbf{0.0066}} & \underline{\textbf{0.0049}} & \underline{\textbf{0.9523}} & \underline{0.9992} \\
         & 2.0 & \underline{\textbf{5}} & \underline{\textbf{0.0045}} & \underline{0.0084} & \underline{0.9523} & \underline{0.9997} \\
         & 4.0 & \underline{\textbf{5}} & \underline{\textbf{0.0005}} & \underline{0.0016} & \underline{\textbf{1.0000}} & \underline{\textbf{0.9999}} \\
         & 8.0 & \underline{\textbf{5}} & \underline{0.0006} & \underline{0.0017} & \underline{\textbf{1.0000}} & \underline{\textbf{0.9999}} \\
        \arrayrulecolor{lightgray}
        \cmidrule[0.5pt](l{1pt}r{1pt}){1-7}

        \multirow{2}{*}[-22pt]{Approx. enum.} & 0.1 & \underline{287} & \underline{\textbf{0.0475}} & 0.0123 & 0.5000 & 0.9995\\
         & 0.3 & \underline{294} & \underline{\textbf{0.0248}} & \underline{\textbf{0.0108}} & \underline{\textbf{0.8261}} & \underline{\textbf{0.9998}} \\
         & 1.0 & \underline{392} & \underline{0.0176} & \underline{0.0058} & \underline{\textbf{0.9523}} & \underline{\textbf{0.9998}} \\
         & 2.0 & \underline{354} & \underline{0.0047} & \underline{\textbf{0.0080}} & \underline{\textbf{1.0000}} & \underline{\textbf{0.9999}} \\
         & 4.0 & \underline{182} & \underline{0.0006} & \underline{\textbf{0.0015}} & \underline{\textbf{1.0000}} & \underline{\textbf{0.9999}} \\
         & 8.0 & \underline{155} & \underline{\textbf{0.0001}} & \underline{\textbf{0.0007}} & \underline{\textbf{1.0000}} & \underline{\textbf{0.9999}} \\
        \cmidrule[0.5pt](l{1pt}r{1pt}){1-7}

        \multirow{6}{*}[-0pt]{\makecell{Laplace \\ mechanism \\ \small{(baseline)}}} & 0.1 & \multirow{6}{*}[-1pt]{-} & 0.0475 & 0.0279 & 0.1980 & 0.3732 \\
          & 0.3 &  & 0.0475 & 0.0279 & 0.1980 & 0.3732 \\
         & 1.0 &  & 1.3991 & 0.0261 & 0.1980 & 0.5656 \\
         & 2.0 &  & 0.9480 & 0.0247 & 0.2631 & 0.7040 \\
         & 4.0 &  & 0.5064 & 0.0216 & 0.4444 & 0.8299 \\
         & 8.0 &  & 0.2401 & 0.0170 & 0.6451 & 0.9074 \\
        \arrayrulecolor{black}
        \bottomrule[1pt]
        \multicolumn{7}{l}{\small*Best: \underline{\textbf{bold}}, second-best: \underline{underline}.}
    \end{tabular}
    }
    \vspace{-6pt}
\end{table*}

\begin{figure}[t]
\captionsetup[subfloat]{farskip=2pt,captionskip=-2pt}
\centering
        \subfloat[$\epsilon$ = 0.1]{\includegraphics[width=1.0\linewidth]{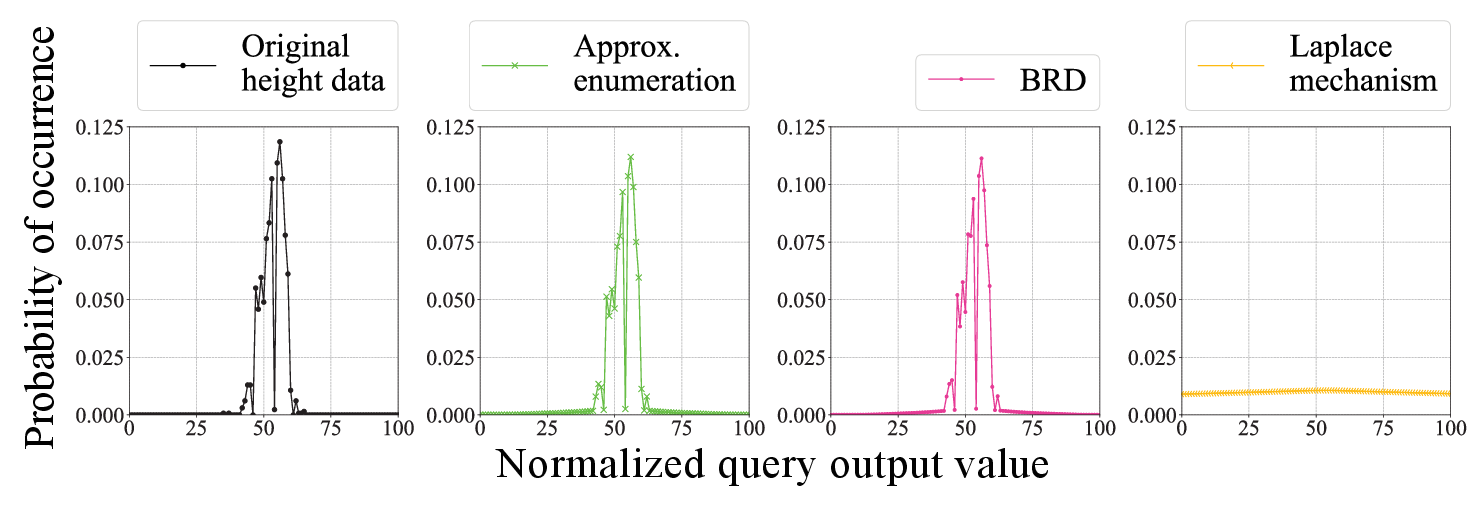}}\\
        \subfloat[$\epsilon$ = 0.3]{\includegraphics[width=1.0\linewidth]{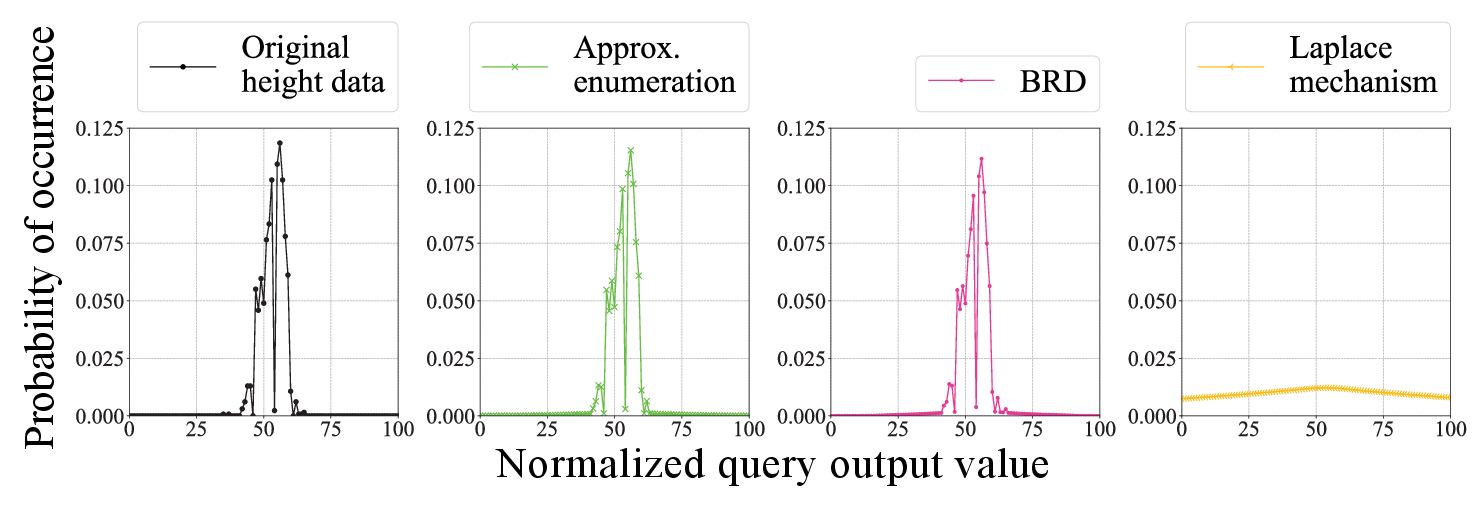}}\\
        \subfloat[$\epsilon$ = 1]{\includegraphics[width=1.0\linewidth]{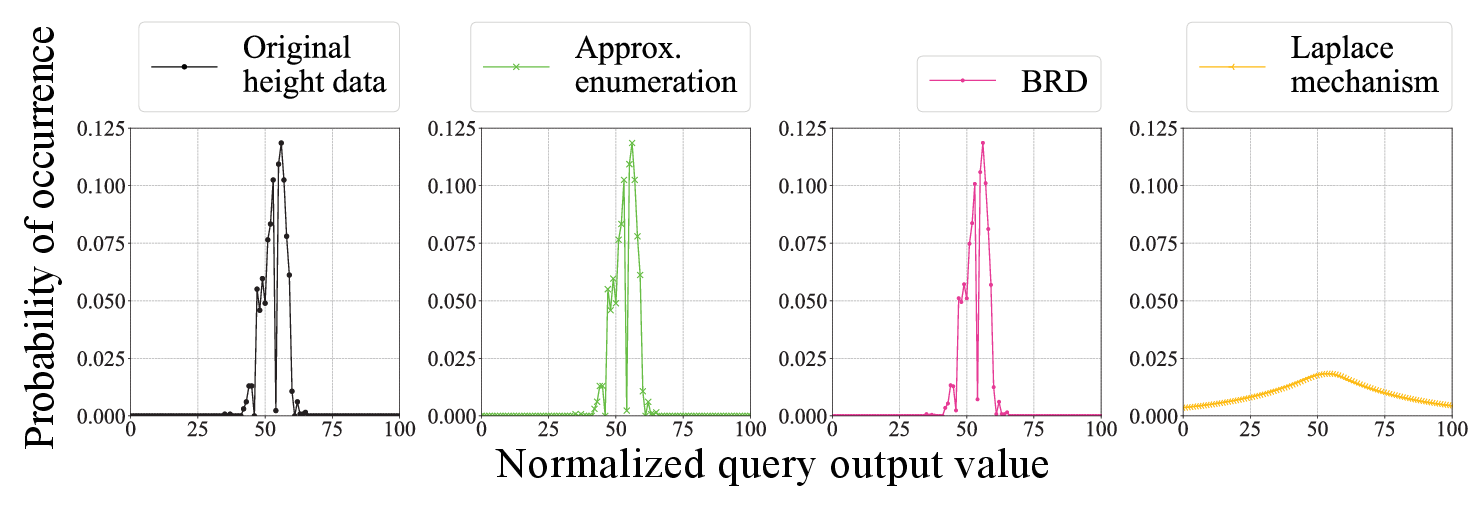}}\\
        \subfloat[$\epsilon$ = 2]{\includegraphics[width=1.0\linewidth]{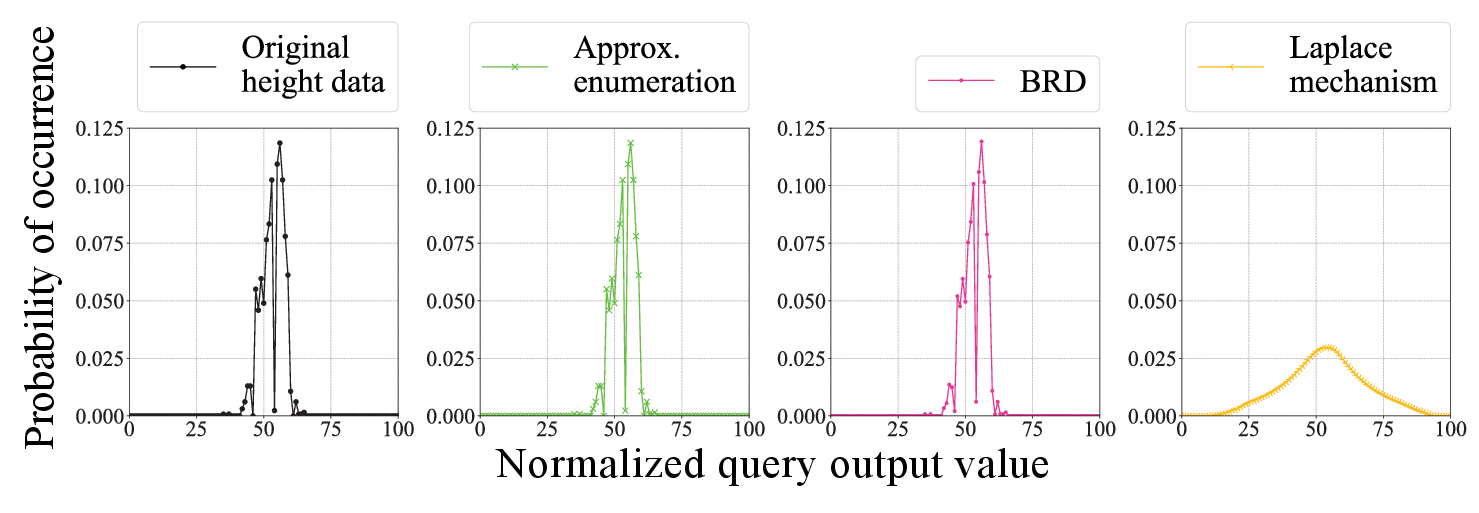}}
\caption{Comparison of query output probability distributions for the height data with each algorithm, when $\epsilon$ = 0.1, 0.3, 1, and 2. The $x$-axis represents the index of the categorization bins from 0 to 100 ($K=101$).}
\label{fig:dist_height}
\end{figure}

\noindent joined five teams from 1963 to 2021: \textit{Atlanta Hawks}, \textit{Boston Celtics}, \textit{Charlotte Hornets}, \textit{Chicago Bulls}, and \textit{Cleveland Cavaliers}.

\paragraph{\textbf{Analysis 1: Preservation of statistical features}.}
Here, we first analyze the preservation of statistical features after executing randomized mechanisms for the height feature of the NBA player dataset. 
In Fig. \ref{fig:dist_height}, we compare the probability distribution of the randomized mechanisms' output. 

As shown in the figure, the proposed NVO game (BRD and AE) has more similar shapes of distribution to the original one than the conventional Laplace mechanism, by executing per-instance noise optimization. 
We can observe that the probability distribution of the Laplace mechanism is getting similar to the original query output distribution as $\epsilon$ increases; however, the proposed NVO game better preserves the shape than with only using $\epsilon=0.1$ than the Laplace mechanism of $\epsilon=8$.

\begin{figure}[t]
\captionsetup[subfloat]{farskip=2pt,captionskip=-2pt}
\centering
        \subfloat[$\epsilon$ = 0.1]{\includegraphics[width=.78\linewidth]{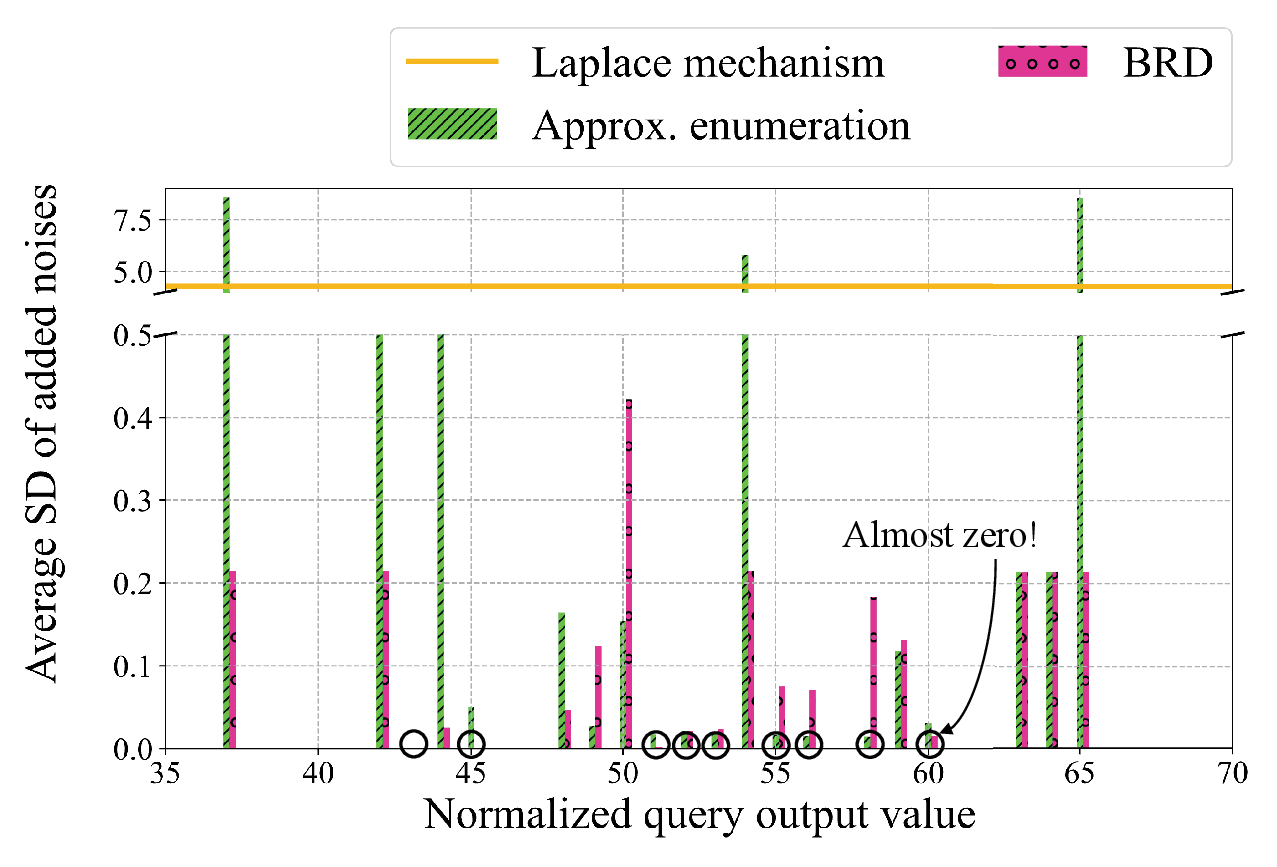}}\\
        \subfloat[$\epsilon$ = 0.3]{\includegraphics[width=.78\linewidth]{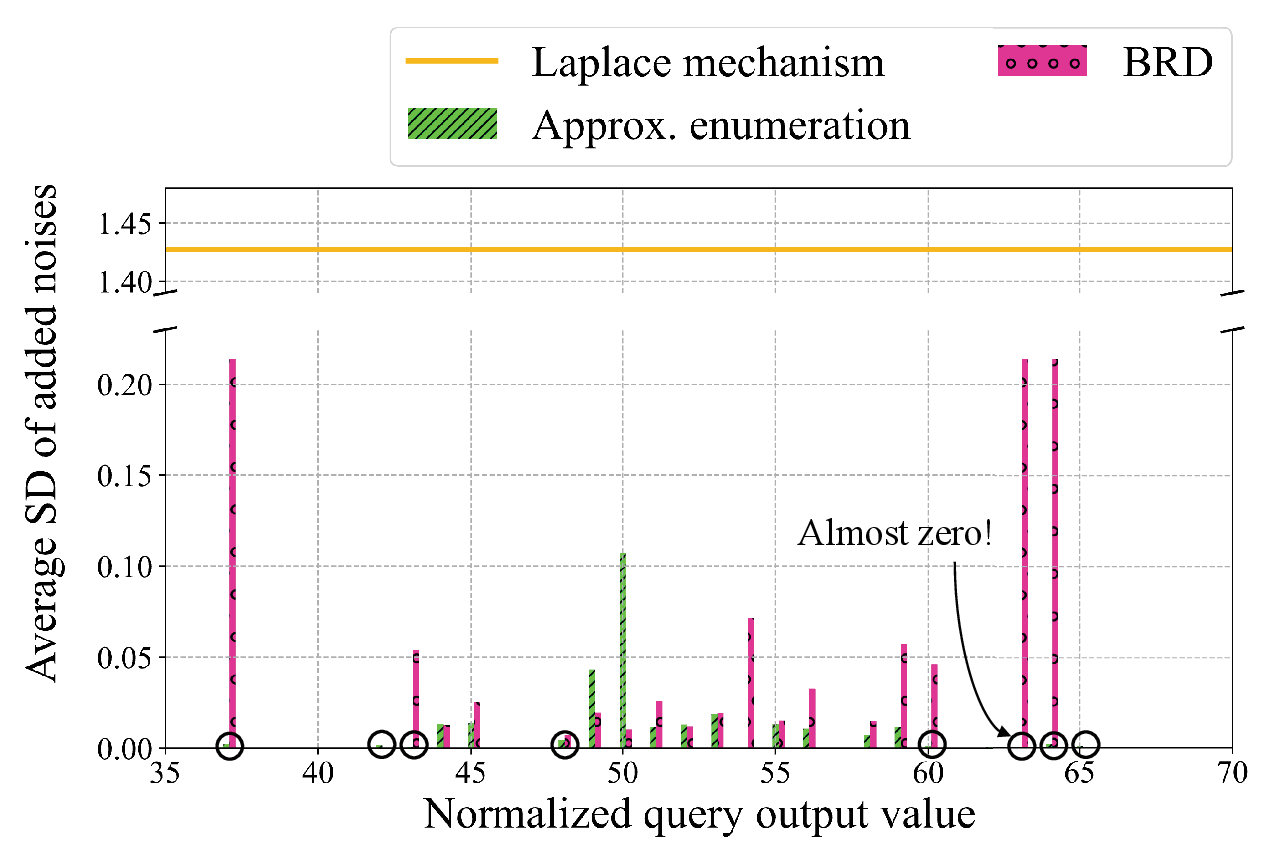}}\\
        \subfloat[$\epsilon$ = 1]{\includegraphics[width=.78\linewidth]{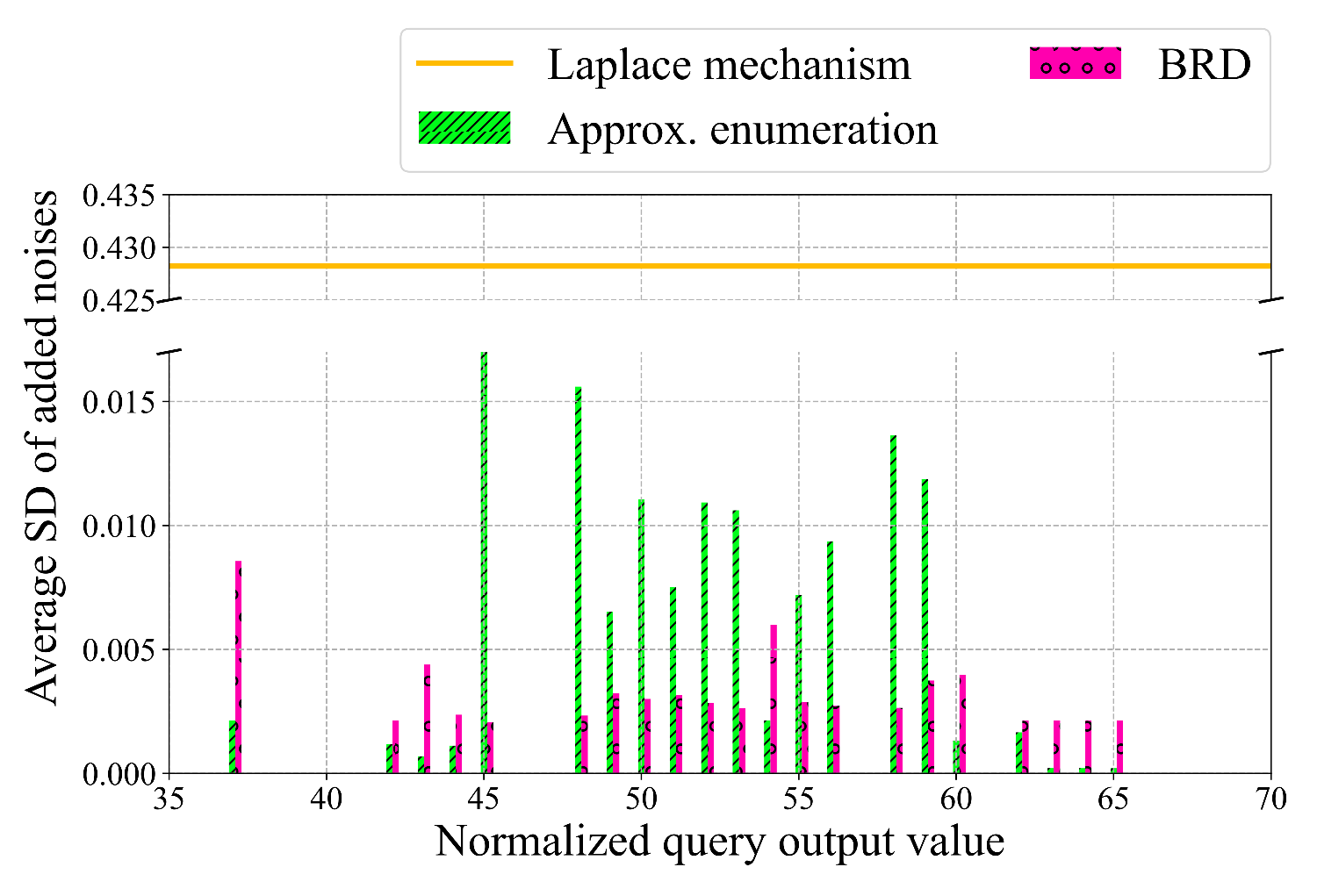}}\\
        \subfloat[$\epsilon$ = 2]{\includegraphics[width=.78\linewidth]{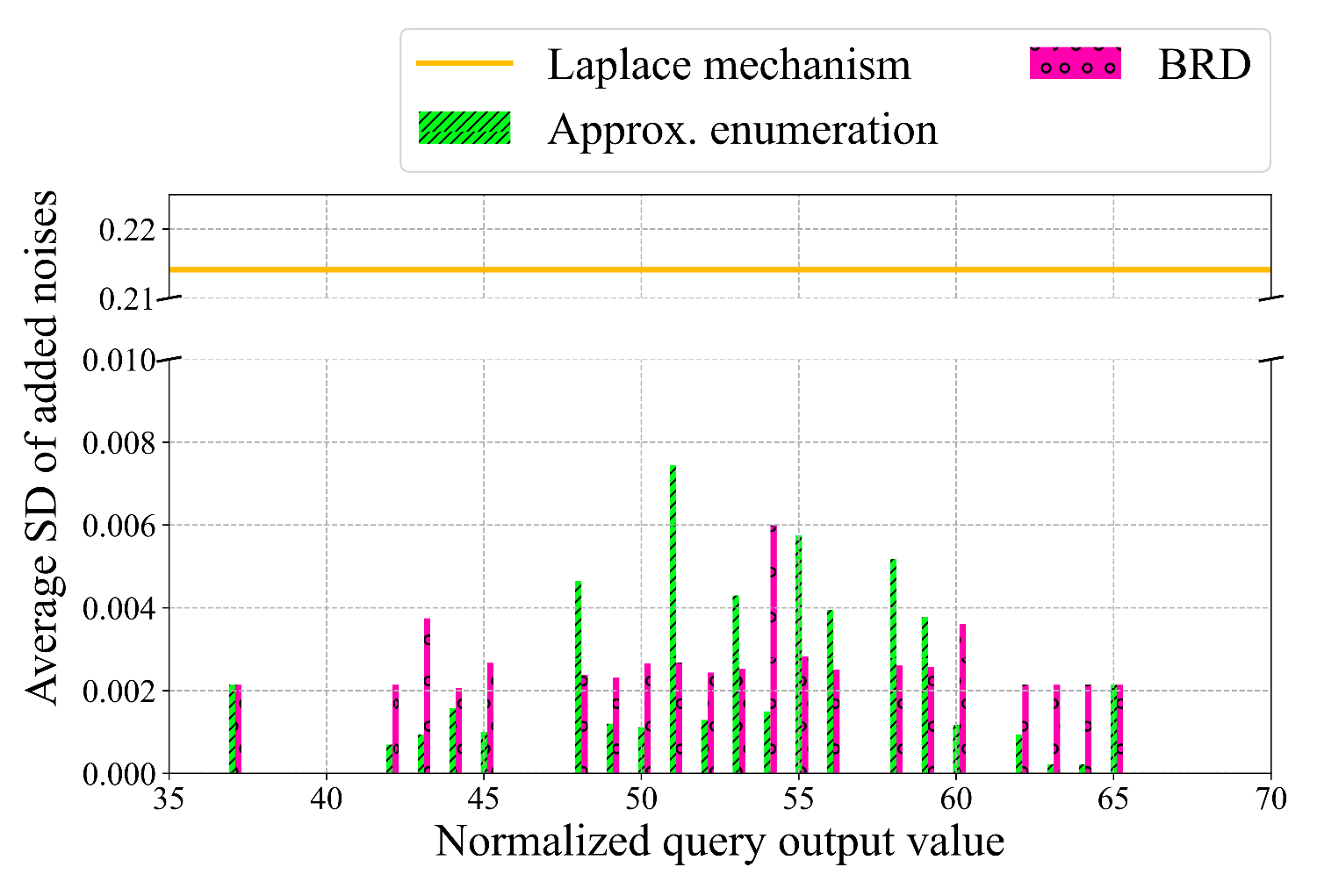}}\\
    \caption{Distributions of average noise standard deviation for the height dataset for $\epsilon$ = 0.1, 0.3, 1, and 2.}
    \label{fig:var_height}
\end{figure}

In Table. \ref{tab:Result_of_Height}, we quantitatively measure various statistical utility functions: 1) KL divergence, 2) L1 loss of standard deviation (SD), 3) Jaccard index, and 4) cosine similarity of the distribution.
In the table, the NVO game-related algorithms have superior statistical utility than the Laplace mechanism at 99.53\%. Despite its superior performance, the AE algorithm requires a much longer computation time than the BRD algorithm.

In Fig. \ref{fig:var_height}, the SD of the added noise in each categorization bin is depicted. 
As we can observe, compared to the conventional Laplace mechanism, the BRD and AE algorithms add relatively small variance noises, thereby achieving superior preservation of statistical features.


\begin{figure}[ht]
\captionsetup[subfloat]{farskip=2pt,captionskip=-2pt}
    \centering
    \subfloat[$\epsilon$ = 0.1]{\includegraphics[width=.5\linewidth]{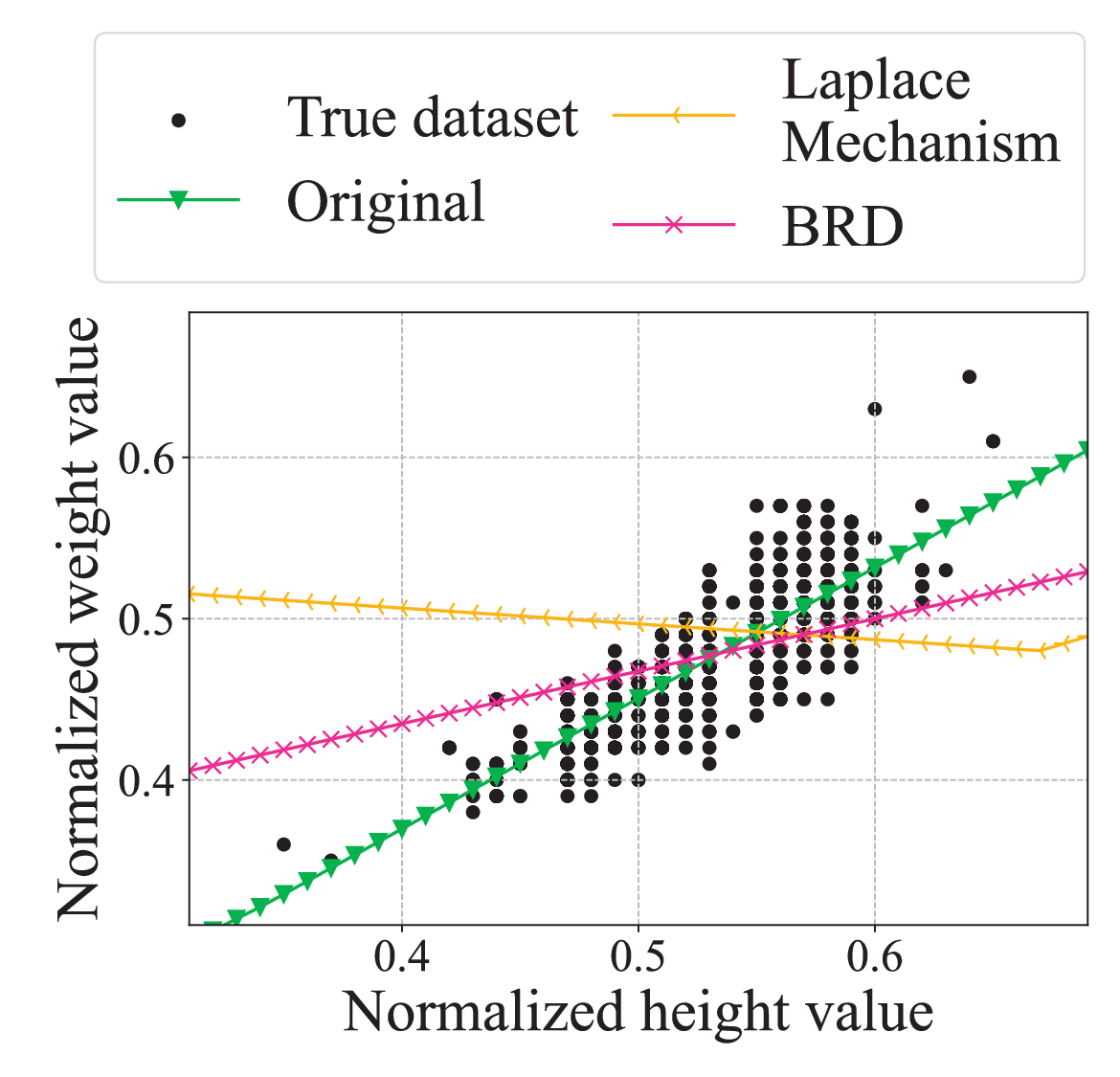}}
    \subfloat[$\epsilon$ = 0.3]{\includegraphics[width=.5\linewidth]{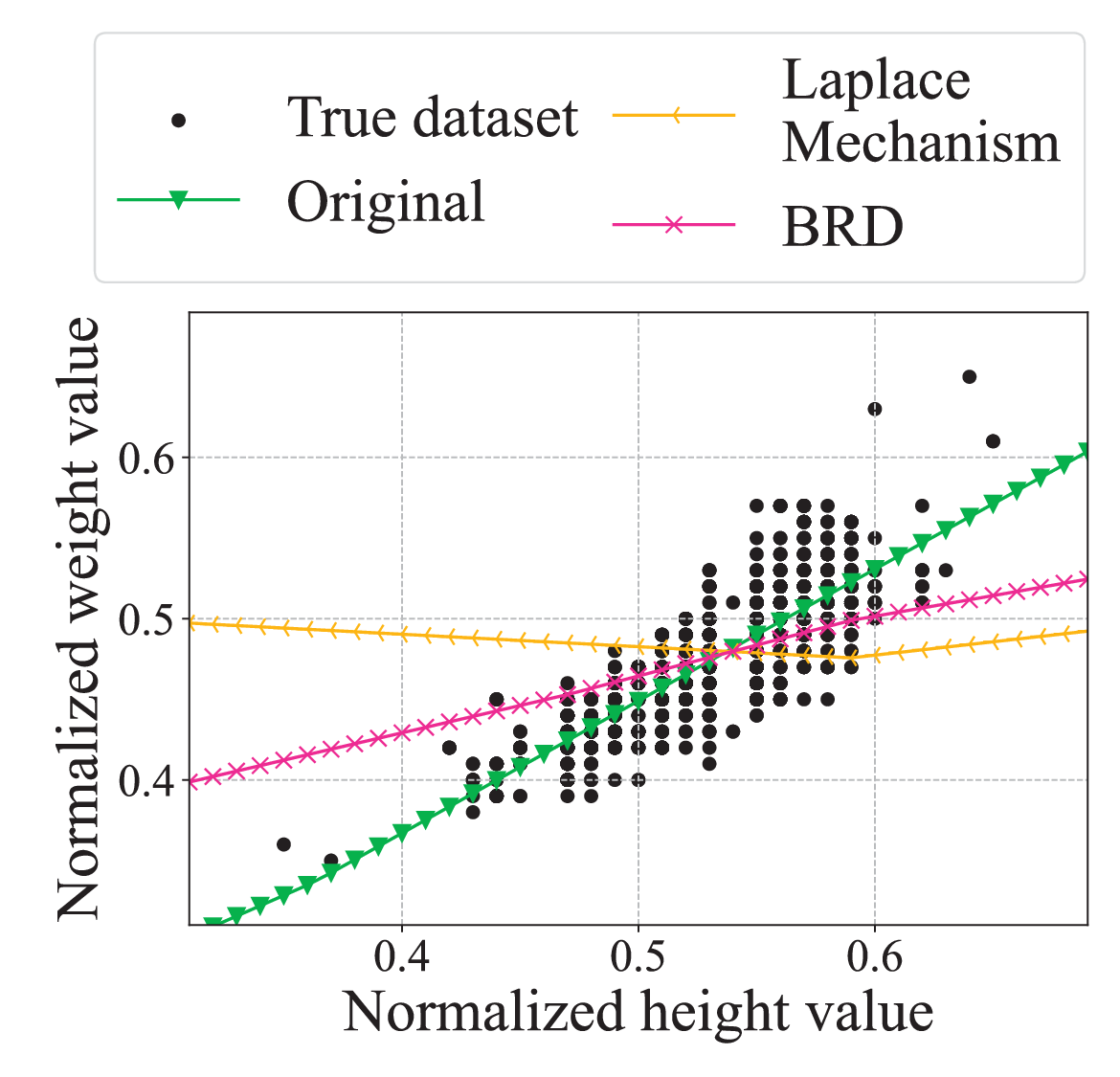}}\\
    \subfloat[$\epsilon$ = 1]{\includegraphics[width=.5\linewidth]{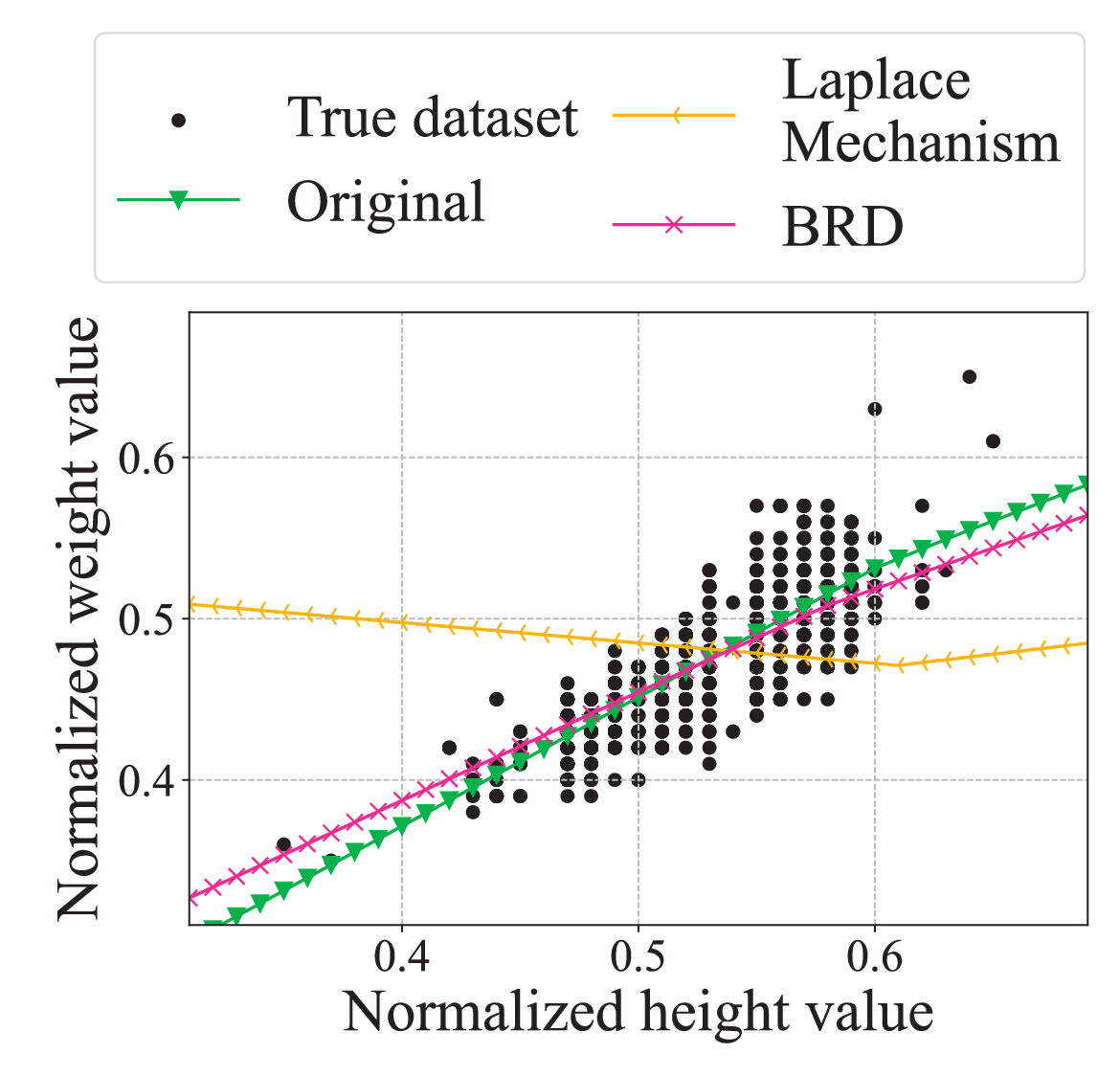}}
    \subfloat[$\epsilon$ = 2]{\includegraphics[width=.5\linewidth]{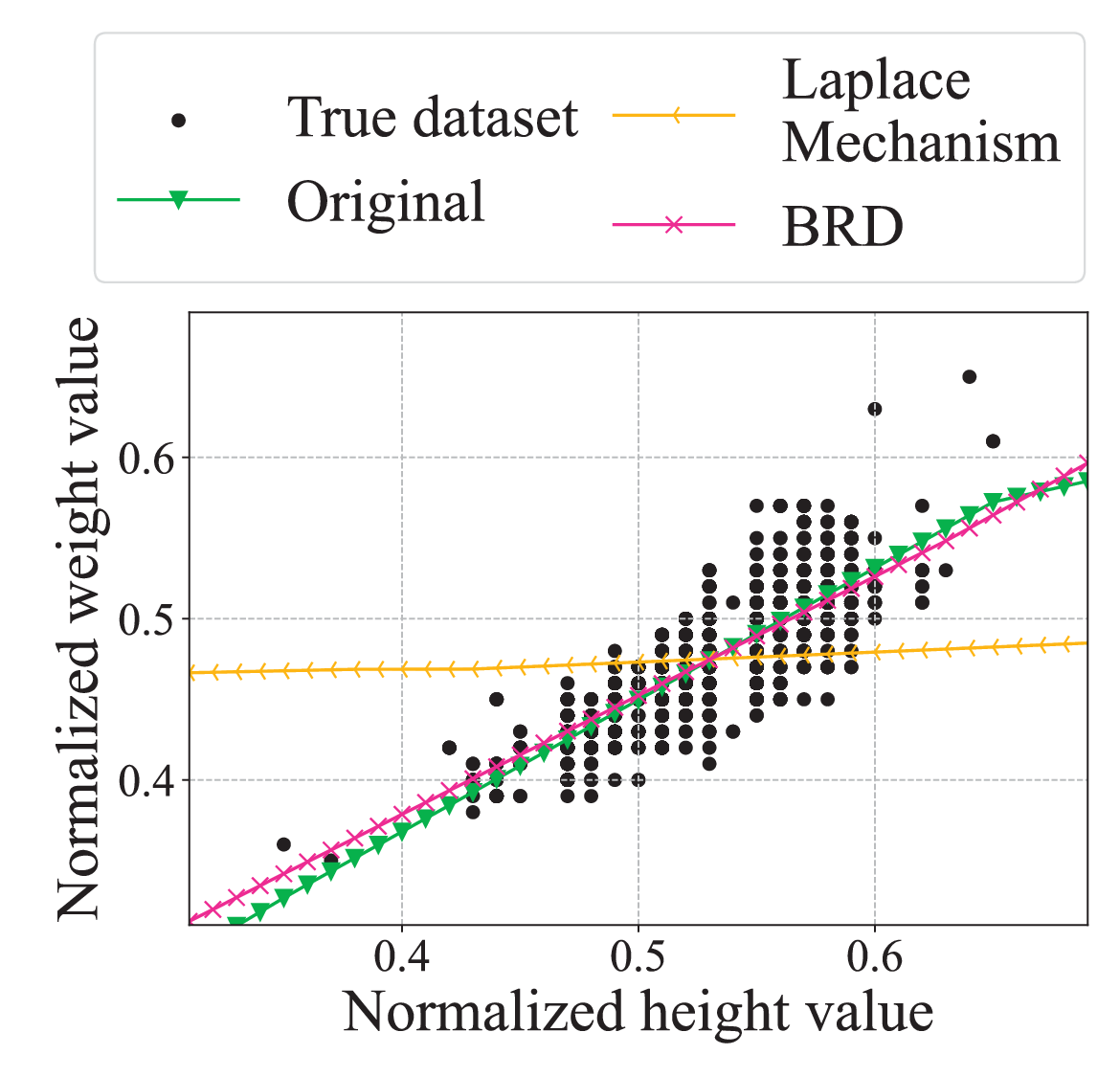}}
    \caption{Linear regression result of 500 sampled data for each algorithm for $\epsilon$ = 0.1, 0.3, 1, and 2.}
    \label{fig:lin_reg}
\end{figure}

 \begin{table}[ht]
\centering
\caption{The average RMSE loss for regression task for the entire dataset, where the samples were generated using the noise associated with the pDP/DP algorithms.}
\label{tab:regression_result}
    \adjustbox{width=1.0\linewidth}
    {
    \begin{tabular}{ccccccc} 
        \toprule[1pt]
        \multirow{2}{*}[-1pt]{Algorithm} & \multicolumn{4}{c}{Average RMSE}\\
         \cmidrule[0.5pt](l{1pt}r{1pt}){2-7}
         & $\epsilon=0.1$ & $\epsilon=0.3$ & $\epsilon=1$ & $\epsilon=2$ & $\epsilon=4$ & $\epsilon=8$ \\
        \cmidrule[0.5pt](l{1pt}r{1pt}){1-7}
        Original data (reference) & \multicolumn{6}{c}{0.0218}\\
        \cmidrule[0.5pt](l{1pt}r{1pt}){1-7}
        NVO game (BRD) & \textbf{0.0278} & \textbf{0.0273} & \textbf{0.0227} & \textbf{0.0221} & \textbf{0.0219} & \textbf{0.0218}\\
        \arrayrulecolor{lightgray}
        \cmidrule[0.5pt](l{1pt}r{1pt}){1-7}
        \makecell{Laplace mechanism} & 0.0516 & 0.0449 & 0.0444 & 0.0380 & 0.0335 & 0.0272 \\
        \arrayrulecolor{black}
        \bottomrule[1pt]
        \multicolumn{5}{l}{\small*Best: \textbf{bold}.}
    \end{tabular}
    }
    \vspace{-3pt}
 \end{table}

\paragraph{\textbf{Analysis 2: Regression task}.}

In order to evaluate the practical usefulness of the randomized mechanisms, we conduct a simulation of a regression task to estimate the weight feature from the height feature of the NBA player dataset. 
For the regression task, we configure a multi-layer neural network, which consists of three layers with ten parameters activated by the Rectified Linear Unit (ReLU) function. 
There are three different neural networks trained with 1) the original dataset, 2) a randomized dataset with the NVO game, and 3) a randomized dataset with the conventional Laplace mechanism.
In Fig.~\ref{fig:lin_reg}, we show the scatter diagram of the preprocessed original dataset, and the height-weight regression curve of the datasets. 
Compared to the conventional Laplace mechanism, the regression curve of the NVO game is more similar to that of the original data.

Quantitatively, as shown in Table \ref{tab:regression_result}, even with the case of low epsilon, 1-DP, the average RMSE of the BRD algorithm is apart from only 8.6\% from that of the original dataset. For 8-DP, the average RMSE for the prediction of the BRD algorithm and original regression are almost the same.

\begin{table*}[t]
\centering
\vspace{-5pt}
\caption{Each algorithm's average computation time, KL divergence, L1 loss of standard deviation, Jaccard index with a threshold of 0.015, and cosine similarity are evaluated for the income data. The modified query output distributions for all algorithms satisfy $\epsilon$-pDP. }
    \adjustbox{width=0.7\linewidth}
    {
\label{tab:Result_of_Income}
    \begin{tabular}{cccc ccc } 
        \toprule[1pt]
        Algorithm & $\epsilon$ & \makecell{Comp. time \\ \tiny{(minutes)}} $\downarrow$ & \makecell{KL \\ divergence} $\downarrow$ & \makecell{L1 SD \\ loss} $\downarrow$ & \makecell{Jaccard index  \\ \tiny{(threshold=0.001)}} $\uparrow$ & \makecell{Cosine \\ similarity } $\uparrow$ \\
        \cmidrule[0.7pt](l{1pt}r{1pt}){1-7}
        \multirow{2}{*}[-11pt]{BRD} & 1.0 & \underline{\textbf{1}} & \underline{0.0223} & \underline{0.0068} & \underline{\textbf{0.9565}} & \underline{0.9997} \\
         & 2.0 & \underline{\textbf{2}} & \underline{0.0087} & \underline{\textbf{0.0003}} & \underline{0.8695} & \underline{\textbf{0.9999}} \\
         & 4.0 & \underline{\textbf{2}} & \underline{0.0014} & \underline{\textbf{0.0003}} & \underline{\textbf{1.0000}} & \underline{\textbf{0.9999}} \\
         & 8.0 & \underline{\textbf{1}} & \underline{0.0013} & \underline{0.0007} & \underline{\textbf{1.0000}} & \underline{0.9999} \\
        \arrayrulecolor{lightgray}
        \cmidrule[0.5pt](l{1pt}r{1pt}){1-7}
        \multirow{2}{*}[-11pt]{Approx. enum.} & 1.0 & \underline{148} & \underline{\textbf{0.0202}} & \underline{\textbf{0.0061}} & \underline{\textbf{0.9565}} & \underline{\textbf{0.9998}} \\
         & 2.0 & \underline{204} & \underline{\textbf{0.0061}} & \underline{0.0026} & \underline{\textbf{1.0000}} & \underline{\textbf{0.9999}} \\
         & 4.0 & \underline{178} & \underline{\textbf{0.0010}} & \underline{0.0017} & \underline{\textbf{1.0000}} & \underline{\textbf{0.9999}} \\
         & 8.0 & \underline{164} & \underline{\textbf{0.0000}} & \underline{\textbf{0.0000}} & \underline{\textbf{1.0000}} & \underline{\textbf{1.0000}} \\
        \cmidrule[0.5pt](l{1pt}r{1pt}){1-7}
        \multirow{4}{*}[-0pt]{\makecell{Laplace \\ mechanism \\ \small{(baseline)}}} & 1.0 & \multirow{4}{*}[-1pt]{-}  & 1.3952 & 0.0274 & 0.2277 & 0.5529 \\
         & 2.0 &   & 0.9373 & 0.0261 & 0.3066 & 0.6959 \\
         & 4.0 &  & 0.4758 & 0.0230 & 0.4782 & 0.8369 \\
         & 8.0 &  & 0.1736 & 0.0178 & 0.6363 & 0.9378 \\
        \arrayrulecolor{black}
        \bottomrule[1pt]
        \multicolumn{7}{l}{\small*Best: \underline{\textbf{bold}}, second-best: \underline{underline}.}
    \end{tabular}
    }
    \vspace{-5pt}
\end{table*}

\begin{figure}[ht]
\captionsetup[subfloat]{farskip=2pt,captionskip=-2pt}
\centering
    \centering
    \subfloat[$\epsilon$ = 1]{\includegraphics[width=.78\linewidth]{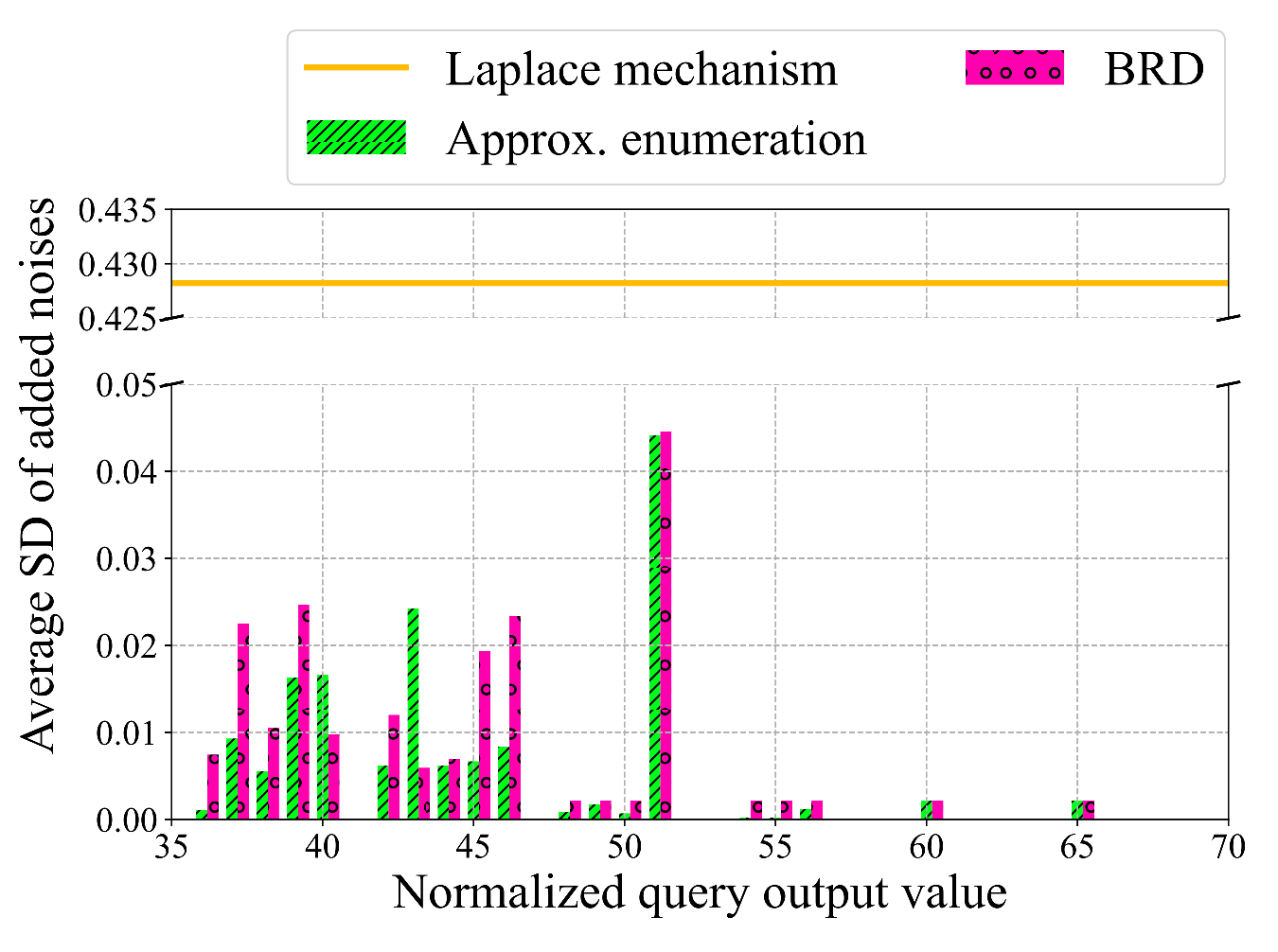}}\\
    \subfloat[$\epsilon$ = 2]{\includegraphics[width=.78\linewidth]{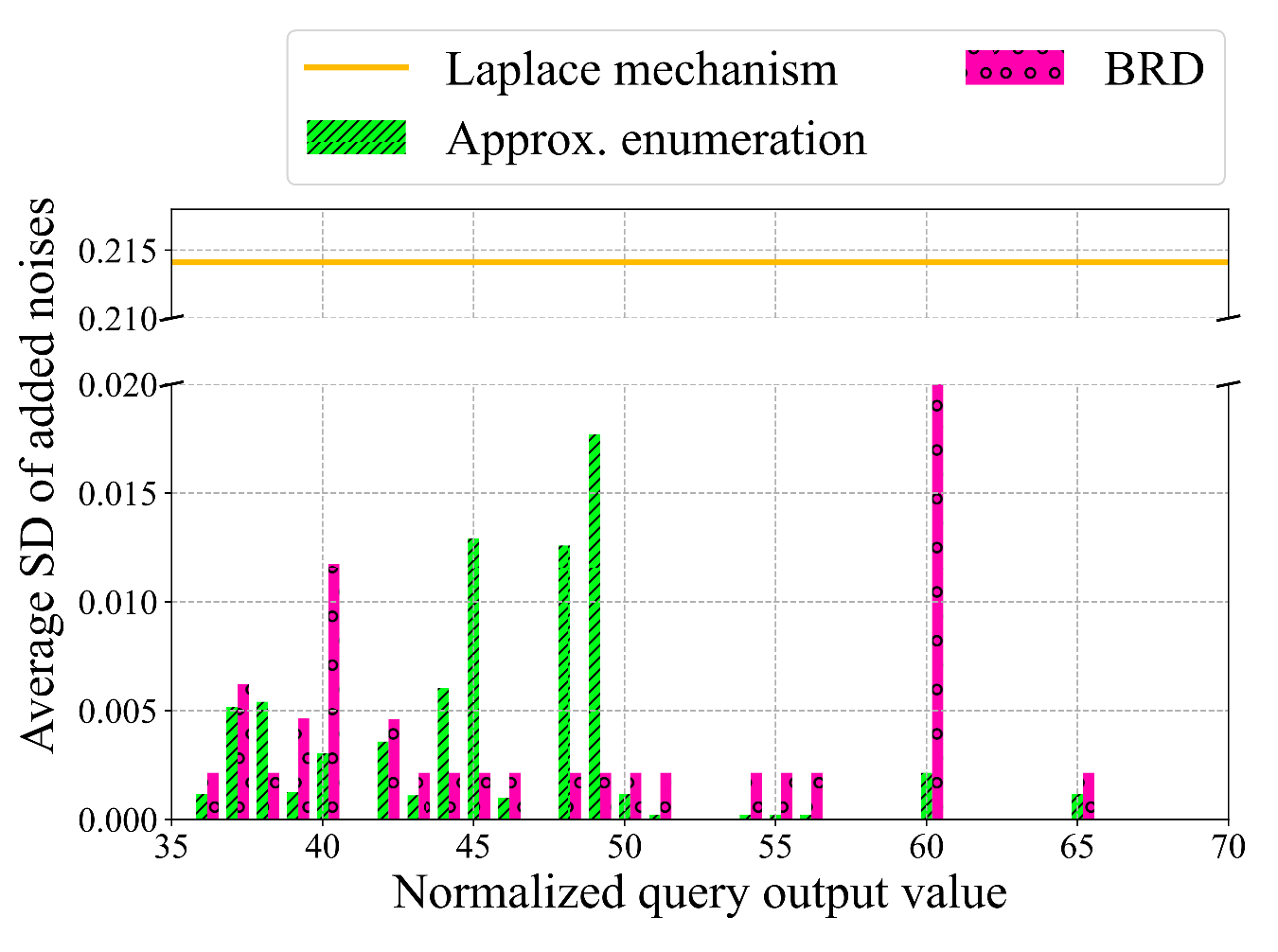}}\\
    \caption{Distributions of average noise standard deviation for the income dataset for $\epsilon$ = 1 and 2.}
    \label{fig:var_income}
\end{figure}

\subsection{Experimental result 2: Income data}

\paragraph{\textbf{Dataset}.} We utilize the test dataset of the personal income dataset, crafted by UC Irvine.
The number of data in the test dataset is 899. 
Similar to the NBA player dataset, the income values belong to one of the 101 categorization bins.
We note that the single feature analysis is conducted here because there is no continuous feature in the dataset except income. 




The qualitative results can be found in Table \ref{tab:Result_of_Income}. The AE algorithm can preserve nearly equivalent data statistics but at a computational cost around 100-150 times higher, involving approximately 270 generations. The BRD algorithm achieves similar performance much more efficiently.
The BRD algorithm achieved up to a 99.71~\% improvement in KL utility than the Laplace mechanism, while guaranteeing 4-DP for every element, on the income dataset.

For the comparison of the per-instance noise variance, we depict the average standard deviation of the added noise in Fig. \ref{fig:var_income}. 
As similar to the results in the NBA player dataset, the proposed NVO game allocates different amounts of noise by considering its probability mass, thereby having better statistical utility. 

From these results, we confirm that the proposed NVO game concisely outperforms the conventional Laplace mechanism.

\section{Conclusion}

In conclusion, our research optimizes noise on a per-instance basis to achieve $\epsilon$-pDP using a Laplace distribution, enhancing statistical utility over traditional methods. This approach is framed as an NVO game, proving that the Nash equilibrium point assures $\epsilon$-pDP for all instances. Our experiments validate that this method significantly outperforms conventional Laplace mechanisms across various utility metrics.
This framework can be universally applied to all statistical queries under differential privacy, as demonstrated by its extension to random sampling queries.
However, the method has limitations, such as reliance on the Laplace distribution and computational constraints due to large datasets. The choice of noise variances is confined to discrete options, even though we demonstrate convergence to $\epsilon$-pDP.
Future research should focus on optimizing noise variances within discrete spaces, exploring alternative noise distributions, and extending the applicability of NVO games to various domains. Additionally, developing algorithms to manage computational complexities and ensure DP guarantees in environments with limited data diversity and access is essential.


\bibliographystyle{ieeetr}
\bibliography{ref}

\appendices

\section{Proof of Theorem \ref{thm:NE_NVO}}
\label{appendix:thm_proof}
\textbf{Theorem 4.1.}
\textit{Let us define the minimum variance in the set of possible action $\mathcal{B}$ as $b_\textnormal{min}\neq0$. Then, $\epsilon$-pDP upholds if the following condition is satisfied:}
    \begin{equation}
        b_{\min} \ge \frac{1}{\log\left(1+ (|\mathcal{D}|-1) (\exp(\epsilon) -1) / K \right)}.
    \end{equation}

For simplicity of the proof, we tackle the situation for the scalar dataset case of the NVO game, \textit{i.e.,} $d=1$. 
We note that the proof can be extended to the vector version by considering each element separately. 
\subsection{Notations}

In this proof, we use the following notations:

\begin{itemize}
    \item  $b_{i}$: Action of the player $i$, the variance of noise added to data $d_i$.
    \item $\left(b_{i}, b_{-i} \right)$: Set of each players' strategy.
    \item $b^{*} = (b_{i}^{*},b_{-i}^{*})$: An NE point of the NVO game.
    \item $b_{\min} = \arg\min_{b\in\mathcal{B}} b$.
    \item $\mathcal{K}$: The set of possible query output values, s.t. $\mathcal{K} \subset [0,1]$.
    \item $m_{i,x}$: The overall probability mass added to $x \in \mathcal{K}$ by the noise assigned to the $i$-th data instance.
    \item $M_{-i,x} = \sum_{j \in [|\mathcal{D}|] \setminus \{i\}} m_{j,x}$
    \item $v_{\min}, v_{\max}$: The minimum and maximum values of the probability mass $m_{i,x}$ added to $x\in\mathcal{K}$ by additive noise to the $i$-th data instance.
    \item $\Pi_{i}(b_{i}, b_{-i})$: The $i$-th player's payoff for the strategy $\left(b_{i}, b_{-i} \right)$.
    \item $P_{i, \textnormal{E}}$, $P_{i, \textnormal{U}}$: The $i$-th player's $P_{\textnormal{E}}$ and $P_{\textnormal{U}}$ for the strategy.
    \item $\Delta P_{i, \textnormal{E}}$, $\Delta P_{i, \textnormal{U}}$: The change of $P_{\textnormal{E}}$ and $P_{\textnormal{U}}$ values for the $i$-th player, s.t. $\Delta P_{i, \textnormal{E}} = P_{i, \textnormal{E}} - P_{i-1, \textnormal{E}}$ and $\Delta P_{i, \textnormal{U}} = P_{i, \textnormal{U}} - P_{i-1, \textnormal{U}}$.
\end{itemize}

\subsection{Assumption} 
Let us assume that we implement the NVO games with a continuous variance space  $\mathcal{B} = [b_{\min}, \infty)$ for $b_{\min} \neq 0$ and the set of possible query output values $\mathcal{X} = [0,1]$.
We do not add noise with a probability of occurring outside the target range of the integration (\textit{i.e.,} $[0,1]$); thus, the probability density function of the Laplace noise is normalized as in (\ref{eq:proof1_4}).

\subsection{Proof}

\paragraph{The worst case to ensure $\epsilon$-pDP for an data instance.}
For the proof of the theorem, we start with the worst case of the $\epsilon$-pDP of an arbitrary data instance.
In order to satisfy $\epsilon$-pDP for an element $d_{i}$, the following condition should be satisfied:
\begin{align}
& \max_{x} \frac{m_{i,x} + M_{-i,x}}{M_{-i,x} \cdot\frac{|\mathcal{D}|}{|\mathcal{D}-1|}} < \max_{x} \frac{m_{i,x} + M_{-i,x}}{M_{-i,x} } \leq \exp (\epsilon) \label{eq:proof1_1}\\
& \Rightarrow
\max_{x} \frac{m_{i,x} + M_{-i,x}}{M_{-i,x}} \leq \max_{x} \frac{m_{i,x} + \min M_{-i,x}}{\min M_{-i,x}} \\ 
& = \max_{x} \frac{m_{i,x} + (|\mathcal{D}|-1)v_{\min}}{(|\mathcal{D}|-1)v_{\min}} \label{eq:proof1_2} \\
& = \frac{m_{i,q(d_{i})} + (|\mathcal{D}|-1)v_{\min}}{(|\mathcal{D}|-1)v_{\min}}
\leq \exp (\epsilon), \label{eq:proof1_3}
\end{align}
where (\ref{eq:proof1_1}) is initialized from the definition of $\epsilon$-pDP.

\paragraph{Find the ${v_{\min}}$.}
The minimum value of the $m_{i,x}$, represented by $v_{\min}$ is obtained by 
\begin{align}
& v_{\min} = \min_{i,x} m_{i,x} = \min_{\overset{\mu,x \in [0,1]}{b\geq b_{\min}}} \frac{ \frac{1}{2b} \exp(-\frac{|x-\mu|}{b}) }{ \int_{0}^{1} \frac{1}{2b} \exp(-\frac{|t-\mu|}{b}) dt }  \label{eq:proof1_4}\\
& = \min_{\overset{\mu,x \in [0,1]}{b\geq b_{\min}}} \frac{ \frac{1}{2b} \exp(-\frac{|x-\mu|}{b}) }{ 1 - \frac{1}{2}\exp(\frac{\mu-1}{b}) - \frac{1}{2}\exp(\frac{-\mu}{b}) } \\
& = \min_{\overset{\mu,x \in [0,1]}{b\geq b_{\min}}} V(\mu, b, x). \label{eq:proof1_5}
\end{align}

In (\ref{eq:proof1_4}), the definition of $v_{\min}$ is rewritten by the Laplace distribution $f(x|\mu, b) = \frac{1}{2b} \exp(-\frac{|x-\mu|}{b})$. For brevity, in (\ref{eq:proof1_5}), we newly define a function $V(\mu, b, x)$.

Then, our focus is to find a value of $\mu$ for the $v_{\min}$, and check the critical points with following conditions:
\begin{align}
& \frac{\partial V}{\partial \mu} = 0 \label{eq:proof1_6}\\
& \Rightarrow  \frac{1}{2b^{2}}\exp(\frac{\mu-x}{b}) \left[ 1 - \frac{1}{2}\exp(\frac{\mu-1}{b}) + \frac{1}{2}\exp(\frac{-\mu}{b}) \right] \\ 
& ~~~~ - \frac{1}{2b} \exp(\frac{\mu-x}{b}) \left[ -\frac{1}{2b} \exp(\frac{\mu-1}{b}) + \frac{1}{2b} \exp(\frac{-\mu}{b}) \right] \label{eq:proof1_7} \\
& = \frac{1}{2b^{2}} = 0,\label{eq:proof1_8}
\end{align}
where (\ref{eq:proof1_7}) holds because of the Laplace distribution's symmetry, thereby making us consider $x \geq \mu$.
Then, from the result of (\ref{eq:proof1_8}), we confirm that there is no critical point that makes $\partial V / \partial \mu=0$.
Also, when $x \geq \mu$, we confirm that the sign of $\partial V / \partial \mu$ is always positive; thus, the minimizer $\mu$ of the function $V(\mu, b, x)$ is zero as follows:
\begin{align}
& \textnormal{sign}\left( \frac{\partial V}{\partial \mu} \right) = \textnormal{sign} \left( \frac{1}{2b} \right) = \frac{1}{2b^{2}} \geq \frac{1}{2b_{\min}^{2}} > 0 \label{eq:proof1_9} \\
& \Rightarrow \arg\min_{\mu} V(\mu, b, x) = 0. \label{eq:proof1_10}
\end{align}


Then, by substituting $\mu=0$ into $V(\mu, b, x)$, we confirm that the minimizer $x$ of the function is one as follows:
\begin{align}
& \frac{\partial V}{\partial x} = \frac{-\frac{1}{b^{2}}\exp(-\frac{x}{b}) }{ 1-\exp(-\frac{1}{b})} < 0 \Rightarrow \arg\min_{x \in [0,1]} V(0, b, x) = 1. \label{eq:proof1_11}
\end{align}


Up to here, we obtained the minimizers $\mu=0$ and $x=1$. By substituting the minimizers, we can obtain the minimizer $b$ as
\begin{align}
& \arg\min_{b \geq b_{\min}} V(0, b, 1) = \arg\min_{b \geq b_{\min}} \frac{1}{\exp(\frac{1}{b})-1} \\
& = \arg\max_{b \geq b_{\min}} \exp(\frac{1}{b}) = b_{\min} \label{eq:proof1_12}\\
&\therefore v_{\min} = \frac{1}{\exp(\frac{1}{b_{\min}})-1}. \label{eq:proof1_13}
\end{align}

\paragraph{Substitute the obtained  ${v_{\min}}$ for getting the worst case.}

From the result of (\ref{eq:proof1_13}) and (\ref{eq:proof1_3}), we have the bound of $\epsilon$, which always guarantee $\epsilon$-pDP. 
Here, we assume the case the $i$-th player does his best to guarantee $\epsilon$-pDP and choose $b_{i}=\infty$.
Then, we have 
\begin{align}
&  \frac{\min \left( m_{i,q(d_{i})} \right) + \frac{|\mathcal{D}|-1}{\exp(1/b_{\min})-1}}{\frac{|\mathcal{D}|-1}{\exp(\frac{1}{b_{\min}})-1}} = \frac{1 + \frac{|\mathcal{D}|-1}{\exp(1/b_{\min})-1}}{\frac{|\mathcal{D}|-1}{\exp(1/b_{\min})-1}} \\ 
& \leq \exp (\epsilon) \label{eq:proof1_14}\\
& \therefore \epsilon \geq \ln \left( \frac{1 + \frac{|\mathcal{D}|-1}{\exp(1/b_{\min})-1}}{\frac{|\mathcal{D}|-1}{\exp(1/b_{\min})-1}} \right), \label{eq:proof1_15}
\end{align}
which can be equivalently written by
\begin{equation}
            b_{\min} \ge \frac{1}{\log\left(1+ (|\mathcal{D}|-1) (\exp(\epsilon) -1) \right)}. \label{eq:proof1_16}
\end{equation}
In (\ref{eq:proof1_15}), we have $m_{i,q(d_{i})} \geq 1$, where the equality holds when $b_{i}=\infty$ and the PDF is uniform.
Finally, there always exists at least one choice to improve the DP guarantee payoff for all elements.

\paragraph{The strategy is improved to finally guarantee $\mathbf{\epsilon}$-pDP for all elements.}
Before the strategy set satisfies the $\epsilon$-pDP for all elements, we have
\begin{align}
& \min_{b_{i}} \Delta P_{i, \textnormal{E}} \geq 1 > \max_{b_i} \Delta P_{i, \textnormal{U}}. \label{eq:proof1_17}
\end{align}
(\ref{eq:proof1_17}) proves that there exists at least a choice to improve the $\epsilon$-pDP guarantee for an element, when the $\epsilon$ is bounded like (\ref{eq:proof1_15}), and by the definition of $P_{\textnormal{U}}$.
Therefore, players choose a strategy to improve $\epsilon$-pDP until guaranteeing for all elements.\\

\paragraph{The Nash equilibrium ensures $\epsilon$-pDP for all elements.}
Assume that the Nash equilibrium point $\left( b_{i}^{*}, b_{-i}^{*} \right)$ does not satisfy the $\epsilon$-pDP for all elements,
\begin{align}
& \Pi_{i}(b_{i}^{*}, b_{-i}^{*}) \geq \Pi_{i}(b_{i}, b_{-i}^{*}) \\
&  \Rightarrow |\mathcal{D}| >  \Pi_{i}(b_{i}^{*}, b_{-i}^{*}) \geq \max \Pi_{i}(b_{i}, b_{-i}^{*}) = \max_{b_{i}} \left( P_{i, \textnormal{E}} + P_{i, \textnormal{U}} \right) \label{eq:proof1_18} \\
& = \max_{b_{i}} \left( P_{i-1, \textnormal{E}} + P_{i-1, \textnormal{U}} + \Delta P_{i, \textnormal{E}} + \Delta P_{i, \textnormal{U}} \right) \\
& = P_{i-1, \textnormal{E}} + P_{i-1, \textnormal{U}} + \max_{b_{i}} \left( \Delta P_{i, \textnormal{E}} + \Delta P_{i, \textnormal{U}} \right) \label{eq:proof1_19}\\
& \geq P_{i-1, \textnormal{E}} + P_{i-1, \textnormal{U}} +1 \geq P_{i-2, \textnormal{E}} + P_{i-2, \textnormal{U}} +2 \geq \dots \nonumber \\ 
& \geq \min_{i, (b_{i},b_{-i})} \left( P_{i, \textnormal{E}} + P_{i, \textnormal{U}} \right) + |\mathcal{D}| = |\mathcal{D}| ,\label{eq:proof1_20}
\end{align}
where (\ref{eq:proof1_18}) follows the definition of Nash equilibrium and the definition of NVO game's payoff.
Because the result in Equations \ref{eq:proof1_17} to \ref{eq:proof1_20} contradicts ($|\mathcal{D}|>|\mathcal{D}|$), we show that the assumption in this paragraph is false. That is, the Nash equilibrium point $\left( b_{i}^{*}, b_{-i}^{*} \right)$ must satisfy the $\epsilon$-pDP for all elements.

\section{Approximated Enumeration for NVO game Via Genetic Algorithm}
\label{Appen:AE}

\begin{figure}[ht]
    \centering
    \includegraphics[width=0.9\linewidth]{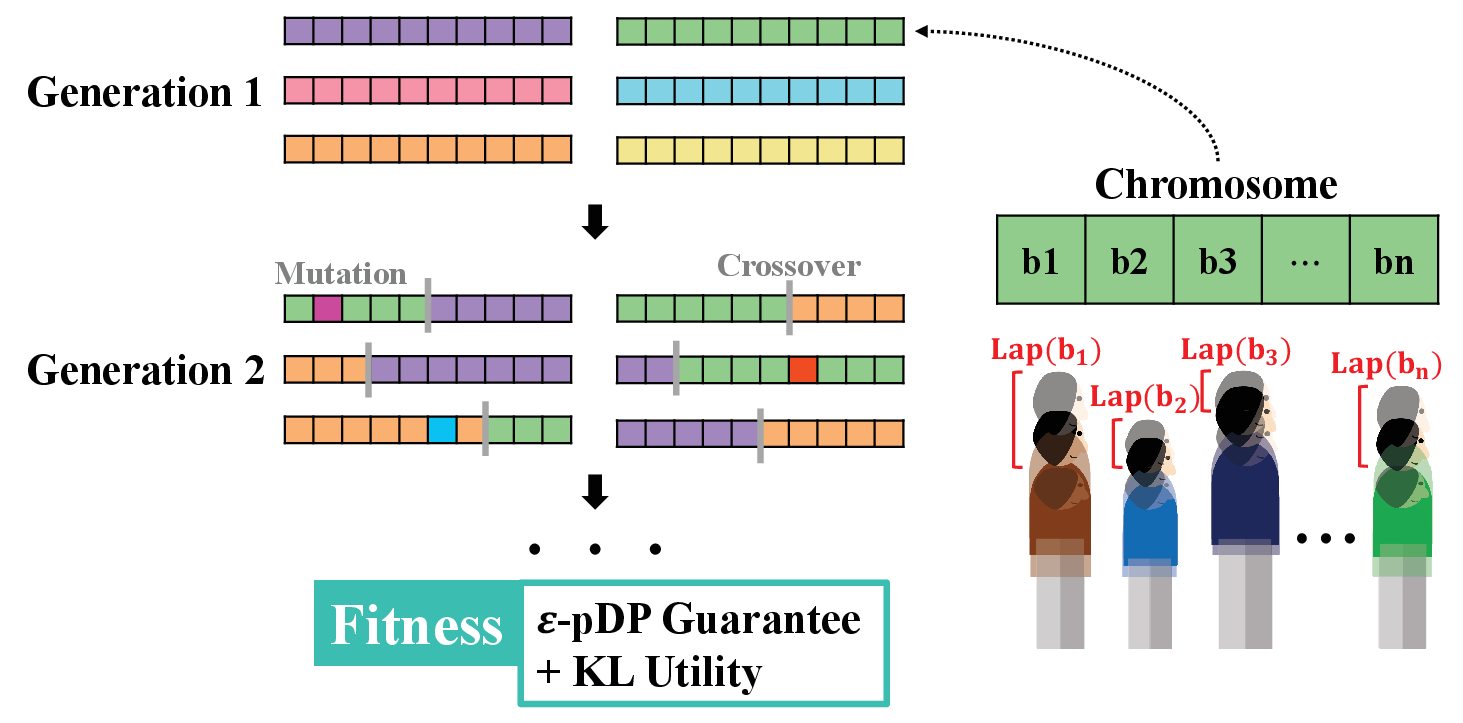}
    \caption{AE for the NVO game via genetic algorithm to find an NE point.}
    \label{fig:GA}
    \vspace{-3pt}
\end{figure}

Enumerating the proposed game precisely proved to be computationally challenging. Instead of that, we adopted an approach grounded in evolutionary game theory. We conducted an approximated enumeration (AE) algorithm by running simulations across numerous generations. This technique monitors strategy evolution over time, revealing promising approaches without the need for an exhaustive exploration of every possible option.

\paragraph{Chromosome.} 
Chromosomes typically symbolize solutions to the specific optimization challenge being addressed. In the framework of the NVO games, each gene is representative of the variance variable $\mathbf{b}_i$ tied to the noise introduced to the query output for every sequential element.

\paragraph{Fitness function.} 
Fitness function serves as the criterion for selecting the most suitable chromosomes that fulfill the specified criteria and can pass down their traits to offspring. 
Hence, we adopt the payoff $P(\mathcal{M}_\textnormal{I}, \mathcal{D})$ as our fitness function.

The initial generation's chromosomes are created by randomly selecting values within the variance space $\mathcal{V}$ for each gene. A larger population broadens the solution search space, minimizing the risk of local optima. Some high-fitness parents are retained in the offspring generation to avoid local optima.
During offspring generation, random crossover points are used, and their optimal number can be determined via simulation. Mutation probability is set to balance between avoiding local optima and ensuring trait transfer.
If the best fitness value remains constant across generations, it suggests an NE approximation. The current chromosome may be optimal, but due to randomness, other solutions might emerge.

With ample time, the AE algorithm has the potential to match the performance of the exact enumeration algorithm and attain an NE point. Our experiments continued for an extended period to ensure convergence. Nevertheless, there is no theoretical guarantee that an NE point is achievable within polynomial time.

\paragraph{\textbf{Hyperparameters}.}
Our proposed BRD algorithm does not require specific hyperparameter settings.
In the AE, we initially set the number of chromosomes in the population to 500, and for each generation, we involve 10 chromosomes in the mating process. We randomly designate 2 crossover points, and we introduce a 5~\% probability for each gene to undergo mutation. We employ a steady-state selection approach, retaining the top 5 parents with the highest fitness values for the next generation. We utilized PyGad~\citep{GAD2021} library for the implementation.

\paragraph{\textbf{Hardware environment}.}
We conduct experiments using an AMD Ryzen Threadripper 1920X 12-Core Processor and 32 GB of RAM. Since there is no need for extensive parallel computations, GPU utilization is not required. To conserve computing resources and facilitate a fair comparison in terms of execution time on the same evaluation criteria, we exclusively relied on CPU computations.

\section{Additional Experimental Results}

\subsection{Experimental result 3: Large Income data}
Essentially, individual privacy is harder to guarantee with smaller datasets due to the increased significance of a data point. Hence, opting for smaller datasets makes privacy assurance more challenging. As datasets grow larger, ensuring privacy becomes comparatively easier. For these reasons, we conduct experiments on a dataset of approximately 1000 in size, but to demonstrate scalability, we also perform experiments on a dataset ten times larger, comprising 10,000 instances.

\paragraph{\textbf{Dataset}.}
We utilize the test dataset of the credit profile dataset.
We conduct our experiments by randomly sampling a cohort of ten thousand individuals. We note the correlation between age and income, and we exploit those two features in our experiments. Similar to the NBA player dataset, the income and age values belong to one of the 101 categorization bins.

\begin{figure}[h]
\captionsetup[subfloat]{farskip=2pt,captionskip=-4pt}
    \begin{minipage}[c]{1.0\linewidth}      
            \centering
            \subfloat[$\epsilon$ = 1]{\includegraphics[width=1.0\linewidth]{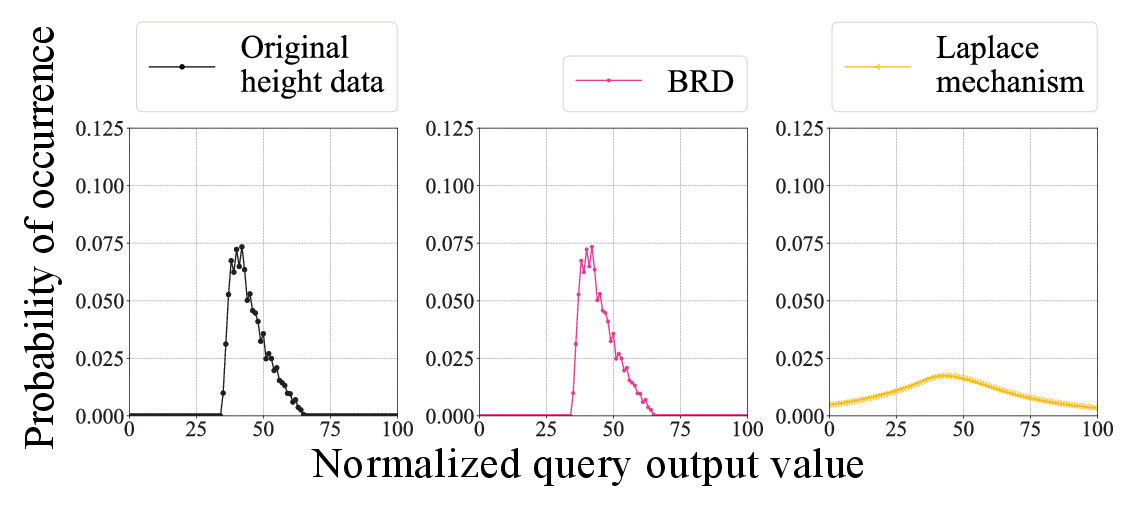}}
            \\
            \subfloat[$\epsilon$ = 2]{\includegraphics[width=1.0\linewidth]{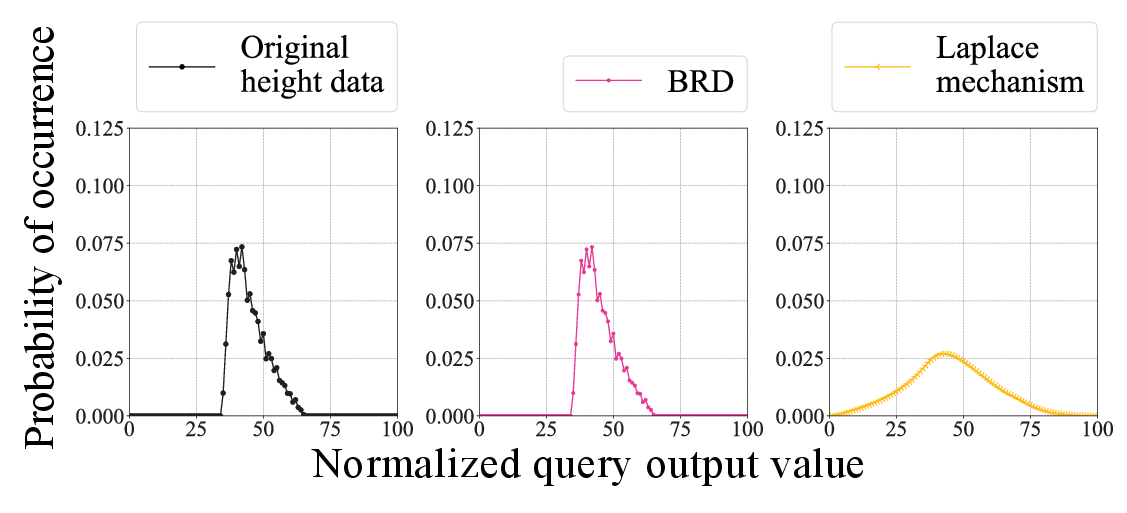}}
    \end{minipage}
    \vspace{-7pt}
\caption{Comparison of query output probability distributions for the large income data with each algorithm, when $\epsilon$ = 1 and 2.}
\label{fig:dist_large_supp}
\vspace{-5pt}
\end{figure}

\paragraph{\textbf{Analysis 1: Preservation of statistical features}.}

In Fig. $\ref{fig:dist_large_supp}$, the random sampling query output of the original data, conventional Laplace mechanism, and NVO game (BRD) is depicted. As similar to the result of the main manuscript, the NVO game better preserves the probability distribution than the conventional Laplace mechanism, by executing per-instance noise optimization. As the dataset size increased, the AE took an excessively long time to converge, preventing us from confirming its convergence within a reasonable timeframe. Consequently, we omitted its results from our findings.

\begin{table*}[t]
\centering
\caption{Each algorithm's average computation time, KL divergence, L1 loss of standard deviation, Jaccard index with a threshold 0.001, and cosine similarity are evaluated for the large income data, for $\epsilon$ = 1 and 2. The modified query output distributions for all algorithms satisfy $\epsilon$-pDP. }
\label{tab:Result_of_Large_supp}
    \adjustbox{width=0.8\linewidth}
    {
    \begin{tabular}{cccc ccc } 
        \toprule[1pt]
        Algorithm & Target $\epsilon$ & \makecell{Comp. time $\downarrow$ \\ \small{(minutes)}} & KL divergence $\downarrow$ & L1 SD loss $\downarrow$ & \makecell{Jaccard index $\uparrow$ \\ \small{(threshold=0.001)}} & Cos. similarity $\uparrow$\\
        \cmidrule[0.7pt](l{1pt}r{1pt}){5-7}
        \cmidrule[0.5pt](l{1pt}r{1pt}){1-7}
        \multirow{2}{*}[-1pt]{BRD} &  $\epsilon$=1 & \textbf{331} & \textbf{0.0000} & \textbf{0.0000} & \textbf{1.0000} & \textbf{1.0000} \\
         & $\epsilon$=2 & \textbf{633} & \textbf{0.0001} & \textbf{0.0004} & \textbf{1.0000} & \textbf{1.0000} \\
        \arrayrulecolor{lightgray}
        \cmidrule[0.5pt](l{1pt}r{1pt}){1-7}
        \multirow{2}{*}[-1pt]{\makecell{Laplace mechanism \\ \small{(baseline)}}} & $\epsilon$=1 & \multirow{2}{*}[-1pt]{-} & 0.9490 & 0.0190 & 0.3069 & 0.6876 \\
          & $\epsilon$=2 &   & 0.5668 & 0.0173 & 0.3333 & 0.8224 \\
        \arrayrulecolor{black}
        \bottomrule[1pt]
        \multicolumn{7}{l}{\small*Best: \textbf{bold}.}
    \end{tabular}
    }
    \vspace{-10pt}
\end{table*}


\begin{figure}[t]
\captionsetup[subfloat]{farskip=2pt,captionskip=-2pt}
\vspace{-4pt}
    \centering
    \centering
    \subfloat[$\epsilon$ = 1]{\includegraphics[width=0.77\linewidth]{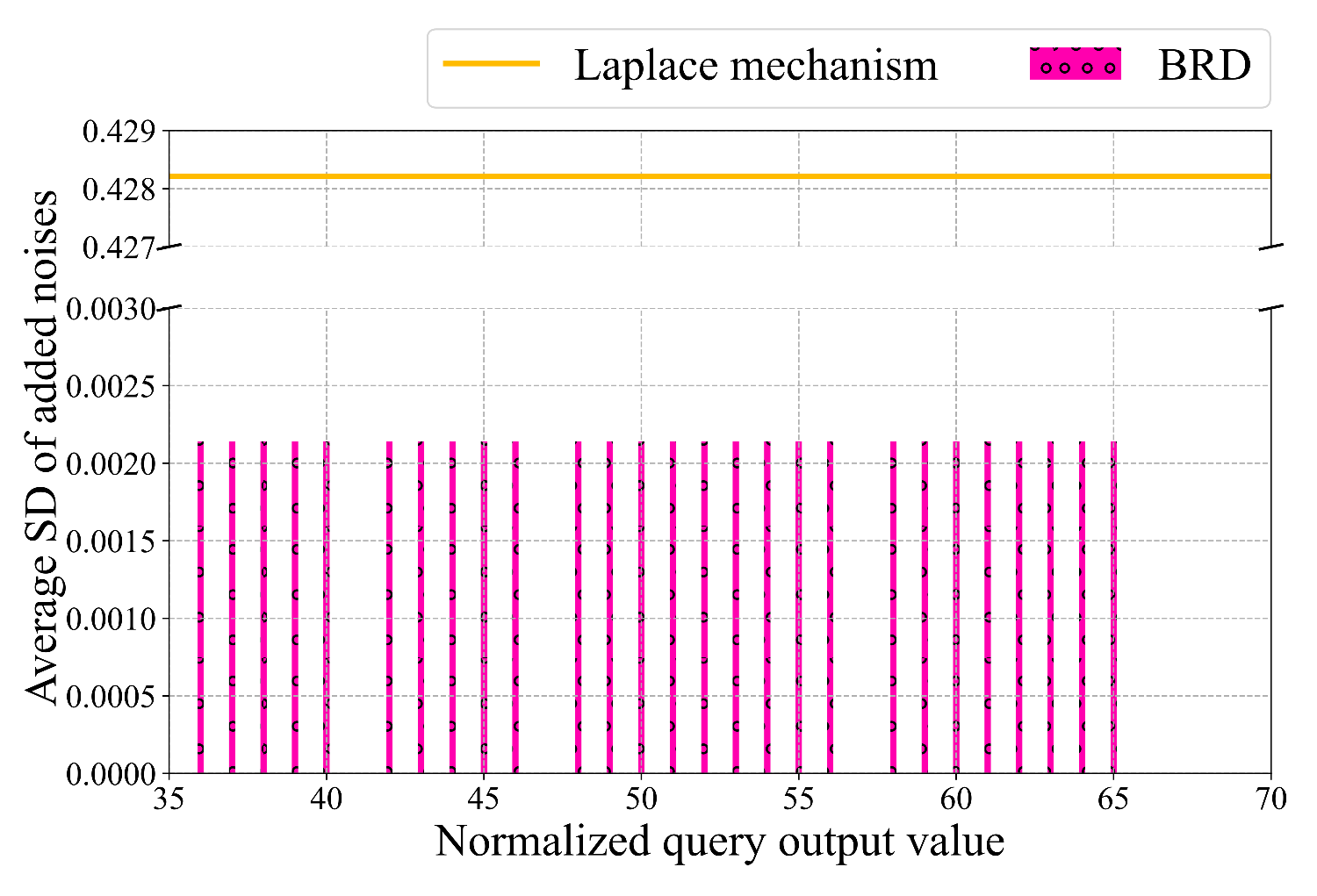}}\\
    \centering
    \subfloat[$\epsilon$ = 2]{\includegraphics[width=0.77\linewidth]{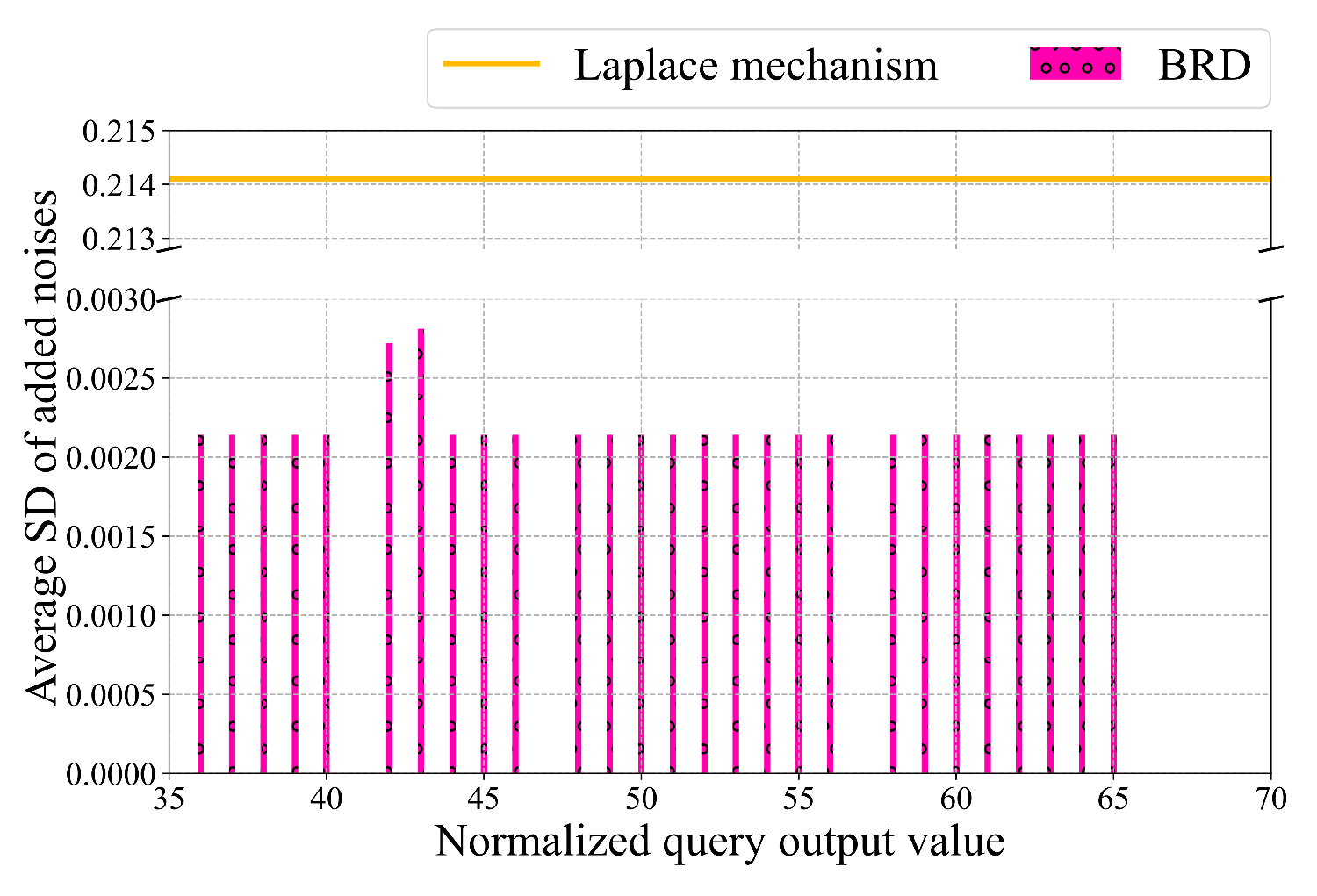}}
    \caption{Distributions of average noise standard deviation for the large income dataset for $\epsilon$ = 1 and 2.}
    \label{fig:var_large_supp}
    \vspace{-7pt}
\end{figure}

In Table $\ref{tab:Result_of_Large_supp}$, the proposed NVO game and the Laplace mechanism are quantitatively evaluated in various statistical metrics. Similar to the results in the main manuscript, the proposed NVO game-based algorithm (BRD) outperforms the Laplace mechanism. Furthermore, for larger datasets, due to the lower individual instance contribution, privacy is better preserved, allowing us to ensure a more robust statistical utility while maintaining the same $\epsilon$-pDP guarantee.

For extremely large datasets, our proposed method incurs high-order computational complexity for the $\epsilon$-pDP guarantee, scaling as $O(|\mathcal{D}|^2)$. To mitigate this, one approach could be to group data points with identical query outputs, allowing for computational reduction through the addition of uniform noise.

In Fig. $\ref{fig:var_large_supp}$, we compare the average standard deviation of the added noise to each categorization bin of the conventional Laplace mechanism and the NVO-game-based algorithm (BRD).
As depicted in the figure, the NVO game adds lower variance at all bins, thereby having better data statistical utility.

\paragraph{\textbf{Analysis 2: Regression task}.}

\begin{figure}[!h]
    \centering
    \vspace{-10pt}
    \begin{minipage}[c]{1.0\linewidth}
        \centering
        \subfloat[$\epsilon$ = 1]{\includegraphics[width=.5\linewidth]{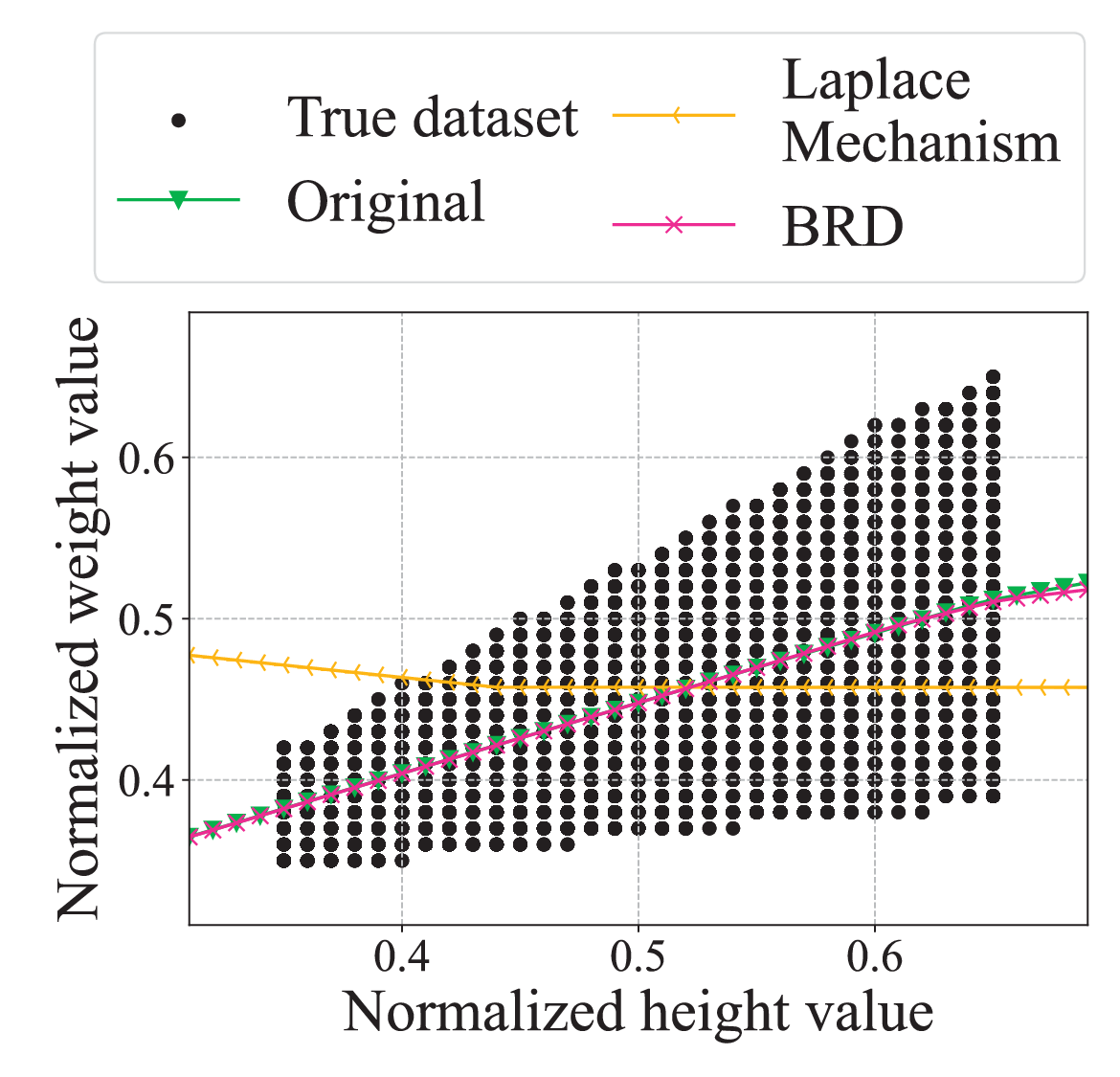}}
        \hfill
        \subfloat[$\epsilon$ = 2]{\includegraphics[width=.5\linewidth]{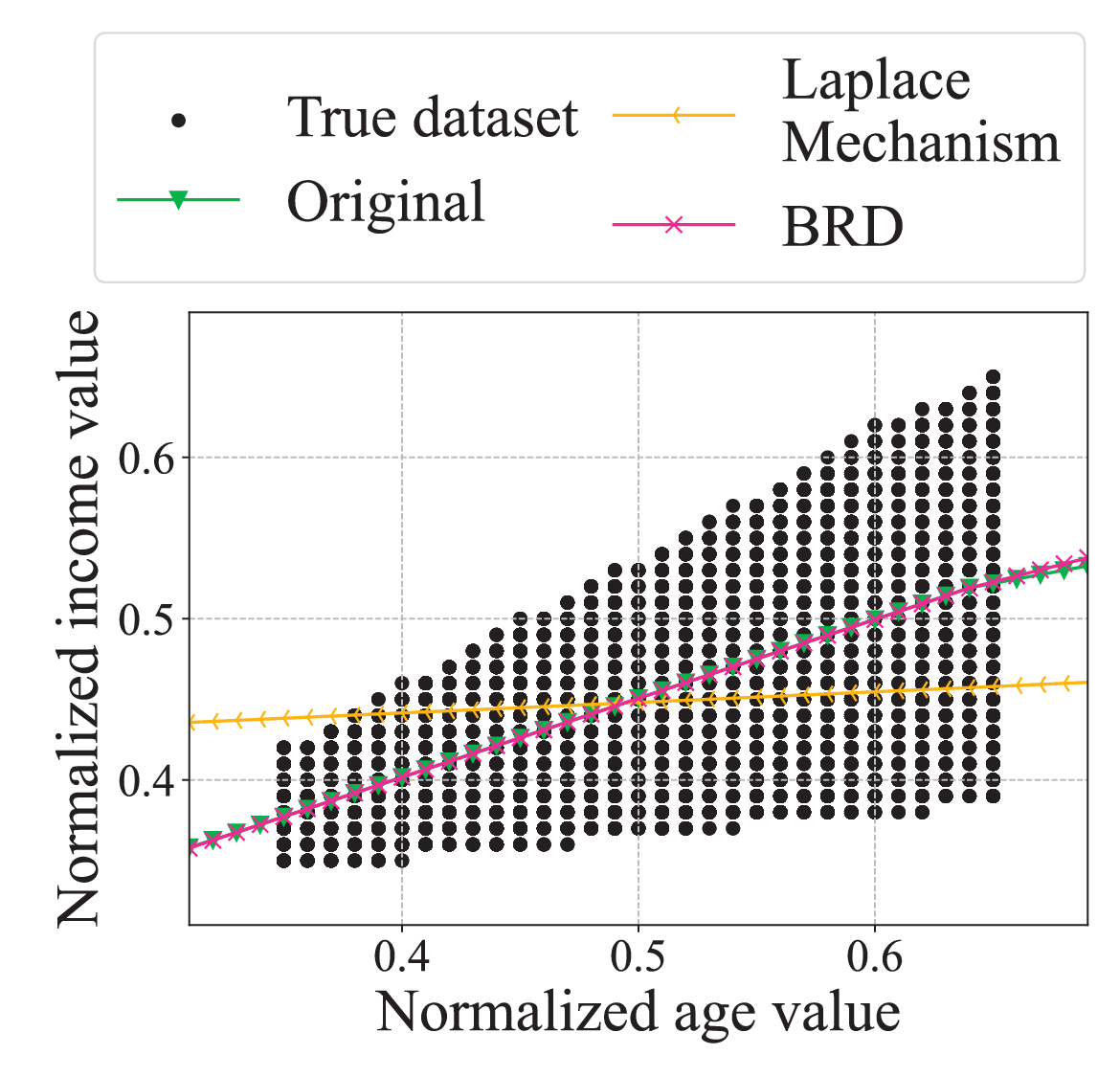}}
    \end{minipage}
    \caption{Linear regression result of 500 sampled data for each algorithm for $\epsilon$ = 1 and 2.}
    \label{fig:lin_reg_large_supp}
\end{figure}

To evaluate the regression task of the proposed NVO game, we run the regression network with three layers. 
Similarly to the main manuscript, the network consists of ten parameters and ReLU activation functions. 
Figure$~\ref{fig:lin_reg_large_supp}$ depicts the scatter diagram of the original dataset and the trained regression line, where the neural network input is age and the output is income. 
For the value of $\epsilon$=1 and 2, the proposed NVO game closely preserves the regression line after applying the randomized algorithm (BRD).
In regression tasks as well, we observe improved data characteristics for the same $\epsilon$-pDP when dealing with larger dataset sizes.

\end{document}